\RequirePackage{xr-hyper}
\documentclass{wlscirep}
\usepackage[utf8]{inputenc}
\usepackage{graphicx}
\usepackage{caption}
\usepackage{xr-hyper}
\usepackage{hyperref}
\usepackage{float}
\usepackage[nameinlink]{cleveref}
\usepackage{xcolor, soul}
\usepackage[normalem]{ulem}
\soulregister\cite7
\soulregister\ref7
\soulregister\Cref7

\makeatletter
\newcommand*{\addFileDependency}[1]{
  \typeout{(#1)}
  \@addtofilelist{#1}
  \IfFileExists{#1}{}{\typeout{No file #1.}}
}
\makeatother

\newcommand*{\myexternaldocument}[1]{
    \externaldocument[SI:]{#1}
    \addFileDependency{#1.tex}
    \addFileDependency{#1.aux}
}

\myexternaldocument{SI}

\crefname{supFig}{Supplementary Figure}{Supplementary Figures}

\graphicspath{{IMGs/}{Figures/}} 

\title{
Unsupervised Tracking of Local and Collective Defects Dynamics in Metals Under Deformation
}

\author[1,$\dag$]{Mattia Perrone}
\author[1,$\dag$]{Matteo Cioni}
\author[1]{Massimo Delle Piane}
\author[1,*]{Giovanni M. Pavan}
\affil[1]{Department of Applied Science and Technology, Politecnico di Torino, Corso Duca degli Abruzzi 24, 10129 Torino, Italy}
\affil[*]{corresponding author: Giovanni M. Pavan (giovanni.pavan@polito.it)}

\affil[$\dag$]{These authors contributed equally.}

\begin{abstract}
Metals owe their unique mechanical properties to how defects emerge and propagate within their crystal structure under stress. However, the mechanisms leading from the early emerging (local) defects to the amplification of dislocations (collective plastic events) are not easy to track. Here, using tensile-stress atomistic simulations of a Copper lattice as a case study, we revisit this classical problem under a new perspective based on local dynamics rather than on purely structural arguments. We use a data-driven approach that allows tracking how local fluctuations emerge and accumulate in the atomic lattice in space and time, anticipating/determining the emergence of local or collective structural defects during deformation. Building solely on the general concepts of local fluctuations and spatiotemporal fluctuation correlations, this approach allows characterizing in a unique way the evolution through the elastic, plastic, and fracture phases, describing metals as complex systems where collective phenomena originate from local dynamical triggering events.
\end{abstract}
\begin{document}
\setstretch{1}
\flushbottom
\maketitle

\thispagestyle{empty}

\section*{Introduction}

Metals represent a fundamental category in materials science, which has been serving as the backbone of technological advancements in the human society from prehistory to modern times.\cite{gordon1988out} Their unique properties, such as high mechanical strength and the ability to deform without breaking (\textit{i.e.}, their prominent plastic phase). These attributes grant metals characteristics, \textit{e.g.}, deformability, weldability, and workability, which make them indispensable for various applications.\cite{sharma2003engineering}

A typical method to assess the mechanical properties of metals is the tensile test, where a traction force is applied to a metal sample to test the materials' response to axial deformation. Such tests typically output characteristic stress-strain ($\sigma - \varepsilon$) curves,\cite{Rivlin1997} where, in the case of metals, two distinct phases can be distinguished: the elastic and plastic phases. In metals, the elastic phase is typically approximately linear, with the proportionality constant between stress and strain defined as the Young's modulus ( $E = \sigma \ \varepsilon$ for $\sigma < \sigma_{y}$, being $\sigma_{y}$ the yield stress, \textit{i.e.}, the limit of the elastic phase).\cite{Chen2016} At the atomic level, this phase involves a reversible stretching of metallic bonds,\cite{Ledbetter1974} where atoms are displaced from their equilibrium lattice positions in the direction of the applied stress, leading to a minimal deformation ($varepsilon$: elongation) of the lattice that fully reversible upon release of the external axial tensile stress $\sigma$. For a higher stress intensity ($\sigma > \sigma_{y}$), metals transition into a non-linear regime defined as the plastic deformation phase, resulting in the irreversible deformation of the material. The primary mechanisms underpinning such plastic lattice deformations are the movement of dislocations – the sliding of crystallographic planes generating linear defects that travel along specific crystallographic directions in the metal's crystalline structure,\cite{anderson2017theory} or either sudden reorganization of the lattice, \textit{i.e.}, twinning.\cite{Christian1995}

Comprehending these mechanisms at the atomic level is crucial to understand the properties if metals, but this requires detecting the emergence of tiny local defects and tracking how these amplify and evolve in collective cascade (catastrophic) phenomena. Tracking these dynamical events at such a detailed atomic level experimentally remains extremely challenging.\cite{cioni2024sampling,ArslanIrmak2021} To address these difficulties, extensive work has been done using molecular models and computer simulations.\cite{Kermode2008} The behavior of metals under deformation has been studied at the quantum level\cite{Linda2023,Kamran2009} and through more approximate models.\cite{MultiscaleMatMod} While quantum treatments offer great detail and accuracy in the treatment of the metal bonds, they struggle in granting a sufficiently large-scale spatial and temporal sampling necessary to avoid finite size effects. The development of (approximated) atomistic force fields (FFs),\cite{Daw1984,Chaturvedi2022} which allows effectively modeling metals and predicting their properties with good accuracy,\cite{RassoulinejadMousavi2016,Wolf2005} offers the opportunity of studying of collective events over large scales and extended time periods.\cite{Kedharnath2021} As notable examples, classical Molecular Dynamics (MD) simulations have been used to study dislocation nucleation, \cite{Nguyen2011,Ryu2011,Li2002,Jennings2013} plastic deformation mechanisms under different loading conditions,\cite{ZepedaRuiz2020,Dezerald2016,An2020} the impact of pre-existing defects,\cite{ZepedaRuiz2017, Singh2011,Saroukhani2016} dislocation interactions,\cite{Zhou1998, Cho2017,Bertin2021, Madec2003,Bertin2024} and even cases of autonomous self-healing.\cite{Xu2013,Xu2016,Barr2023} 

Common simulation approaches for studying metal deformation with MD consist of performing a series of out-of-equilibrium simulations, which can be carried out by applying an external force (controlled stress) or by deforming the simulation box at a controlled strain rate.\cite{Kedharnath2021} These simulations generate MD trajectories that are typically analyzed to investigate the deformation mechanisms by using different types of general parameters, such as coordination numbers or Steinhardt's order parameters,\cite{Lechner2008} or by using tools specifically tailored for studying metals, \textit{e.g.}, centrosymmetry parameters,\cite{Tsuzuki2007} algorithms for identifying dislocations and stacking faults,\cite{Larsen2016,Maras2016,Faken1994} calculation of Burgers vectors\cite{Stukowski2012}(that can identify the collective sliding of crystalline planes), etc. While the use of \textit{ad hoc} parameters specifically tailored for metals may allow to detect and track specific known events, these heavily rely on prior knowledge of the system under study, and may risk losing track of events that are not known \textit{a priori}. Furthermore, different types of descriptors (\textit{e.g.}, centrosymmetry parameters or Burgers vector) are separately used to track different types of (respectively, local or collective) events: \textit{e.g.}, the emergence of points defects or of dislocations. While providing a rich picture of metals, using more agnostic and general approaches (that do not necessarily build on previous knowledge of the system) may be desirable, and may eventually provide new perspectives under which revisiting widely studied phenomena \cite{cioni2023innate, rapetti2023machine, cioni2024sampling}
In addition to these traditional methods, agnostic descriptors, \textit{e.g.}, the Smooth Overlap of Atomic Position (SOAP),\cite{Bartk2013} have been developed to transform the local configuration of atomic neighbors around every atom in a system into spectral representations (high-dimensional fingerprints of structural order) that, capturing information on the radial and angular displacements of atoms, may be useful for characterizing metallic systems.\cite{rapetti2023machine,cioni2023innate}  Despite its advantages, pattern-recognition analyses based on SOAP and parameters capable of capturing dominant structural domains in metals have certain limitations. The emergence of (rare) local defects risks to be typically overlooked (negligible statistical weight) and the permutation invariance of SOAP hinders its ability to capture time-dependent dynamics, crucial for understanding processes, \textit{e.g.}, defect migration and surface diffusion in metals. To overcome these challenges, recent advancements have introduced time-dependent descriptors that focus on tracking even rare/sparse local changes in atomic order over time. Notable examples are, \textit{e.g.}, TimeSOAP\cite{Caruso2023} and the Local Environments and Neighbors Shuffling (LENS) descriptors,\cite{Crippa2023} which are general and abstract descriptors that have been proven effective in capturing subtle, highly-dynamic, and rare local fluctuations in a variety of different systems. LENS, for example, has allowed the capture of rare atomic events that lead to significant structural rearrangements in metallic systems.\cite{Caruso2023, Crippa2023, crippa2023machine} For instance, it has successfully detected dynamic surface changes such as atom detachment and reabsorption on \textit{fcc}-Cu surfaces, and local fluctuations that trigger major structural changes, such as the transformation of vertices into rosette configurations, in Au nanoparticles.\cite{Crippa2023, cioni2024sampling} These tools guided a change of approach: from pattern-recognition analyses of the most prominent structural environments that compose a system, to a systematic detection and correlation of statistically-relevant fluctuations in the (noisy) time-series data obtained from the individual trajectories of all atoms composing the system\cite{becchi2024layer, LEAP}. This results in a general information-rich approach that may be used to study any complex molecular system without prior assumptions.
Here, we use such abstract and general approaches based on fluctuations and fluctuation correlations to explore the internal behaviour of metals under tensile load, and we systematically compare the results that we obtained with those attainable with classical descriptors (\textit{e.g.}, centrosymmetry parameter and Burgers vector). We explore the physical behavior of a metallic system under a constant strain rate, with a focus on its elastic, plastic, and fracture phases. Copper (Cu), a face-centered cubic (\textit{fcc}) metal, is chosen as a prototypical case study due to its well-characterized mechanical properties. Utilizing LENS, an agnostic descriptor of neighbor fluctuations within each atom's local environment, we demonstrate how from the LENS time-series it is possible to clearly distinguish the elastic and the plastic phases during deformation from the characteristic local and collective fluctuations emerging within them. We also extend our analysis to the fracture event and to the post-fracture dynamic behavior in the metal, capturing an inertial, relax phase where atomic dynamics is still present along newly formed (imperfect) ductile-fractured surfaces. Our results demonstrate how, simply building on the abstract concepts of local (LENS) fluctuations, and of their spatial and temporal correlations, it is possible to efficiently capture local phenomena captured by \textit{ad hoc} tailored parameters such as the centrosymmetry parameter, as well as collective behaviors identified by Burgers vectors. This provides a comprehensive picture of the material's response to stress and, in a completely ''agnostic'' data-driven way it allows to reconstruct and track the full sequence of mechanical events — from the emergence of local defects, to the nucleation and amplification dislocations, to the final fracture and the subsequent relax surface dynamics of the new surfaces that are formed. This also offers a new perspective where the mechanical behaviour and properties of metals are revisited by monitoring how the microscopic atomic dynamics emerges, develops and intensifies within the material, rather by focusing on traditional structural and energetic parameters.

\section*{Results and Discussion}
\subsection*{Elastic-plastic Transition in Cu deformation}
\begin{figure}[H]
\centering
\includegraphics[width=\linewidth]{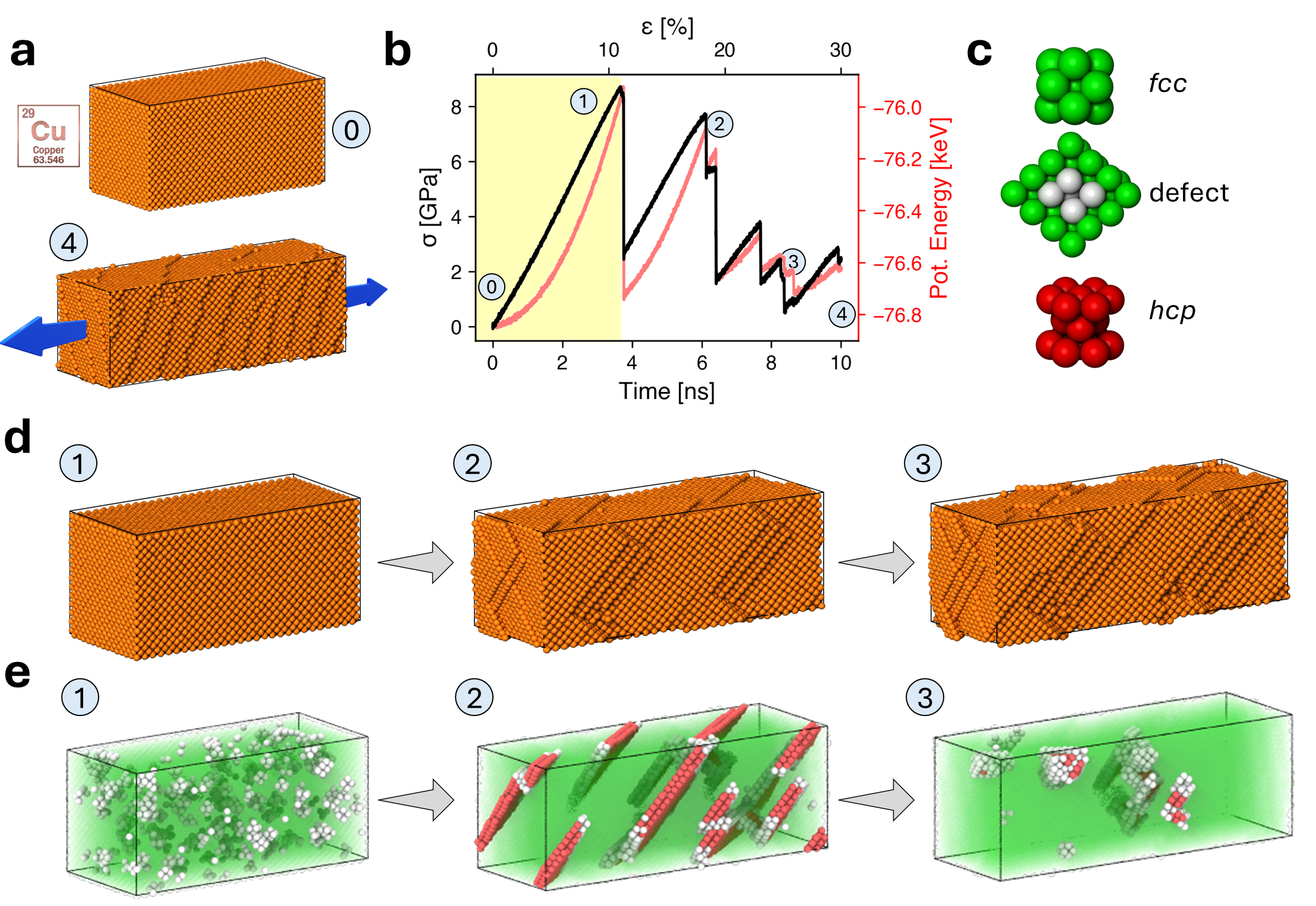}
\caption{Mechanical response of a copper (Cu) supercell under uniaxial stress, captured through molecular dynamics (MD) simulations. 
(a) The initial (0) and deformed (4) states of the copper supercell are depicted, with deformation along the X-axis, as indicated by the blue arrows. 
(b) Stress-strain curve (black) illustrating the mechanical response, transitioning from elastic to plastic deformation, with corresponding changes in potential energy (red). The numbered markers on the curve correspond to the states illustrated in the other panels. Data shown are from one simulation replica (\textit{cf.} Supplementary Information (SI) for the complete dataset).
(c) Classification of atoms based on their local atomic environments: \textit{fcc} (face-centered cubic, green) representing stable regions, defective atoms (gray) indicating disrupted crystalline order, and \textit{hcp} (hexagonal close-packed, red) representing stacking faults.
(d) Structural evolution during deformation as seen in the 3D atomic arrangements at stages labeled in (b), representing initial, intermediate, and deformed states.
(e) Corresponding visualization of defect classification during deformation, with atoms colored as per (c), highlighting the emergence and propagation of stacking faults and other defects.
}
\label{fig:fig1}
\end{figure}

We start by simulating a defect-free Copper (Cu) supercell containing 21,952 Cu atoms at temperature of 300 K.
All the simulations are conducted under periodic boundary conditions, in such a way as to model the bulk of Cu under tensile stress effectively. Following preliminary equilibration, a constant strain rate of \(3 \times 10^{7}\)$s^{-1}$ was applied along the \textit{x} direction. The system was simulated for 10 nanoseconds, at the end of which the deformation was equal to 30\% (Fig.~\ref{fig:fig1}a, \textit{cf.} Methods section for the detailed simulation protocol). To gather reliable statistics and account for potential variability in atomic behavior, we performed a series of 6 independent replica MD simulations. In all the figures shown herein, the time-series plots showing specific quantities as a function of time refer to one representative replica, while corresponding plots for the remaining replicas are provided in the Supplementary Information. All other data and analyses presented are based on aggregated statistics from the full set of replicas.

As illustrated in Fig.~\ref{fig:fig1}b, the stress-strain curve initially shows a linear increase in stress, indicating a linear elastic phase where the copper lattice deforms uniformly under the applied load. This elastic regime is maintained until $\sim$3.5 ns, after which point the material transitions into the plastic phase. This transition is accompanied by a notable drop in potential energy, as indicated by the red curve in Fig.~\ref{fig:fig1}b (secondary y-axis), due to the emergence of dislocations within the material. As strain continues to be applied, the material undergoes work hardening, characterized by multiple stress drops (relaxation) and increases, where nontrivial atomic rearrangements occur.\cite{Buehler2005}
To track the atomic-level structural evolution of the material during deformation, we classified the local environment surrounding each atom in the system, as either \textit{fcc} or \textit{hcp}, based on their geometric arrangement and local symmetry. Atoms that did not conform to such configurations were labeled as ''defects'', representing regions where the crystalline order was disrupted (Fig.~\ref{fig:fig1}c). This structural analysis has been conducted using the dislocation extraction algorithm DXA,\cite{Stukowski2012} that relies on common neighbor analysis for crystal structure identification.\cite{Honeycutt1987}
Fig.~\ref{fig:fig1}d-e show the system configurations during progressive deformation, corresponding to the points highlighted in Fig.~\ref{fig:fig1}b. In particular, Fig.~\ref{fig:fig1}e presents the system with atoms classified as shown in Fig.~\ref{fig:fig1}c. The initial phase (Fig.~\ref{fig:fig1}e, step 1), corresponding to the strained lattice before the emergence of the first dislocation in the elastic regime, is characterized by \textit{fcc} atoms being stretched by the imposed strain, alongside the presence of clusters of ''defected'' atoms. As noted in Refs.\cite{Wo2005,Ngan2006}, these atoms represent short-lived dynamical ''hot spots'', lasting on the order of 10s of femtoseconds, which randomly appear and disappear in the lattice. These configurations do not represent structural defects \textit{per se} but rather momentary asymmetric stretching events of the metallic bonds in the atomic neighborhood of certain atoms in the lattice. As the imposed stress or strain increases, these domains have a growing likelihood of evolving into dislocations.\cite{Ryu2011,Nguyen2011} During the first two potential energy drops, plane slipping occurs along preferential directions (Fig.~\ref{fig:fig1}e, step 2): in an \textit{fcc} crystal, this leads to the emergence of stacking faults, identified by atomic structures in \textit{hcp} packing.\cite{anderson2017theory} Finally, at high levels of deformation ($\ge$20\%), the material evolves through non-trivial interactions between dislocations, resulting in a more disordered, complex ensemble structure (Fig.~\ref{fig:fig1}e, step 3).\cite{Buehler2005,Bulatov2006}
It is important to note that the system analyzed in this study incorporates two key approximations compared to real-world tensile experiments. First, our model assumes an idealized, defect-free lattice, whereas real materials typically exhibit a pre-existing density of dislocations. This simplification enables us to isolate and examine the nucleation of defects in an initially pristine material. Second, the strain rates used in MD simulations significantly exceed experimental ones and represent a necessary compromise due to computational limitations.\cite{Yu2019,Chang2017,Amadou2023,Jasperson2025,Bertin2024,Bertin2023,ZepedaRuiz2017,Stukowski2012elasticplastic} These high rates of deformation are known to suppress thermally activated phenomena, such as dislocation nucleation and propagation, thereby underestimating plastic activity. However, this suppression aligns with our objective to focus on the initial nucleation mechanisms that trigger the onset of plasticity, rather than fully capturing the plastic phase.

\subsection*{Tracking local atomic dynamics \textit{via} CSP and LENS descriptors}

\begin{figure}[H]
\centering
\includegraphics[width=\linewidth]{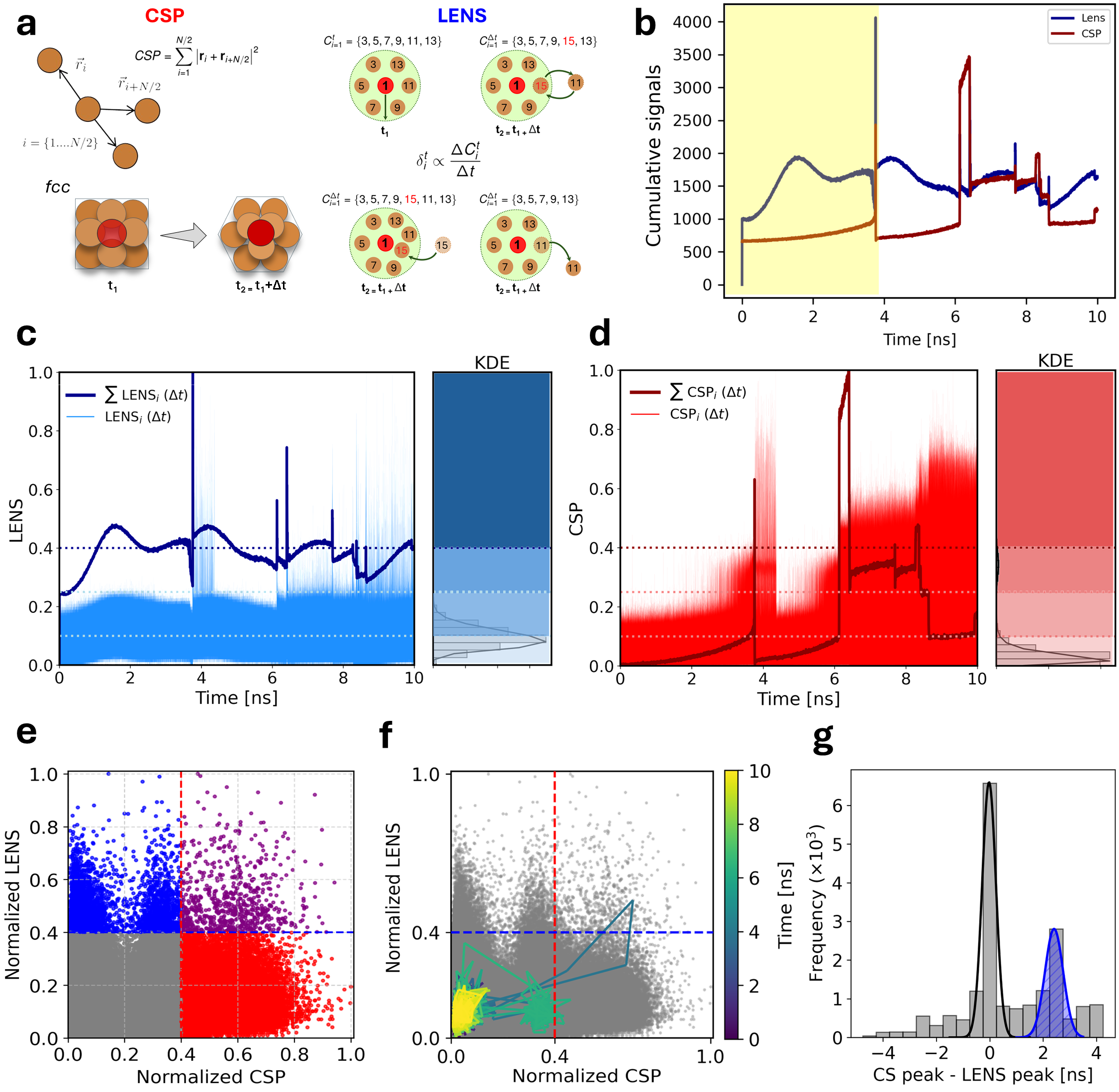}
\caption{
Correlation and temporal evolution of the LENS and CSP descriptors during tensile deformation of Cu.
(a) Schematic representation of the calculation methods for CSP (left) and LENS (right).
(b) Temporal evolution of summed LENS (blue) and CSP (red) values, with the elastic phase highlighted (yellow).
(c, d) Time series of individual atomic LENS (c) and CSP (d) values, with Kernel Density Estimate (KDE) plots on the right showing descriptor value distributions. The summed curves from (b) are overlaid for reference.
(e) Scatter plot of normalized LENS vs. CSP values for each atom across the simulations, with dashed lines dividing the quadrants: low rearrangement/low disorder (gray), low rearrangement/high disorder (red), high rearrangement/low disorder (blue), and high rearrangement/high disorder (purple).
(f) Temporal trajectory of an individual atom in the normalized CSP-LENS plane, color-coded by time.
(g)Histogram of the distribution of time delays between CSP and LENS peak maxima. The black curve represents near-simultaneous peaks, while the blue curve indicates LENS peaks preceding CSP peaks.
}
\label{fig:fig2}
\end{figure}

To attain a more comprehensive analysis of the local atomic environments and their evolution throughout the deformation process, we employed two complementary indicators (Fig.~\ref{fig:fig2}a): the CentroSymmetry Parameter (CSP)\cite{Kelchner1998} and the Local Environments and Neighbors Shuffling (LENS)\cite{Crippa2023} descriptor. CSP was chosen for its ability to quantify deviations from perfect symmetry and to detect local lattice imperfections, making it well-suited for identifying structural defects. In contrast, LENS captures dynamic changes in the local atomic environments that surround each atom in the system over time, allowing us to track subtle, localized rearrangements and fluctuations that may precede larger structural transformations. Together, these indicators provide a framework for capturing both static and dynamic local features of atomic behavior during deformation.

CSP quantifies deviations from a perfectly symmetric environment around each atom: low CSP values indicate a symmetric, defect-free environment typical of an ideal crystal structure, while higher values reveal the presence of asymmetry. The CSP value of an atom, which has \(N\) nearest neighbors (\(N = 12\) for \textit{fcc} and \(N = 8\) for \textit{bcc} lattices), is calculated as:

\begin{equation}
\label{def_CSP}
\text{CSP} = \sum_{i=1}^{N/2} \left| \mathbf{R}_i + \mathbf{R}_{i+N/2} \right|^2,
\end{equation}

where \(\mathbf{R}_i\) and \(\mathbf{R}_{i+N/2}\) are vectors from the central atom to a pair of opposite neighbor atoms. In an ideal centrosymmetric crystal lattice, the contributions from all neighbor pairs cancel out, resulting in a CSP value of zero. However, the CSP becomes positive in a defective crystal region, where the neighborhood is disturbed and non-centrosymmetric. By applying an appropriate threshold to account for minor perturbations due to thermal displacements and elastic strains, CSP can be effectively used as an order parameter to identify atoms that are part of crystal defects.

On the other hand, LENS provides a dynamic, time-dependent quantification on the evolution of local atomic environments by tracking dynamical changes in the identities of neighbor atoms over time\cite{Crippa2023}. Specifically, LENS measures how neighboring atoms are added, removed, or reshuffled between consecutive time intervals, thereby capturing the temporal dynamics of structural rearrangements\cite{Crippa2023} (Fig. \ref{fig:fig2}A).

The LENS value for each atom \textit{i}, denoted as \( \delta_{i}^{t+\Delta t} \), measures these changes between two consecutive time steps \textit{t} and \( t + \Delta t \). It is defined as:

\[
\delta_i^{t+\Delta t} = 
\frac{\#(C_i^{t} \cup C_i^{t+\Delta t} - C_i^{t} \cap C_i^{t+\Delta t})}
{\#(C_i^t + C_i^{t+\Delta t})},
\]

where \( C_{i}^{t} \) represents the list of neighbor atoms within a cutoff distance around atom \( i \) at time \( t \), and \( C_{i}^{t + \Delta t} \) refers to the neighbor list at the next time step.
The numerator counts the number of neighbors who have changed, while the denominator normalizes this value by the total number of neighbors at both time steps.

LENS has proven highly effective in detecting and classifying local fluctuations and dynamic regions that might be overlooked by conventional pattern recognition analyses \cite{Crippa2023,becchi2024layer, cioni2024sampling}. It has been successfully applied to describe the dynamics of metallic systems such as surfaces and nanoparticles, revealing insights that drive phase transitions and structural reorganizations.\cite{Crippa2023, cioni2024sampling,LEAP}

Both descriptors provide atom-level information for characterizing metals, but they differ in important ways. CSP relies on a predefined reference structure to determine the number of neighboring atoms, as shown in Eq.\ref{def_CSP}, making it a system-specific descriptor. In contrast, LENS is agnostic, in that it does not depend on a specific particle type or neighborhood order\cite{Crippa2023}. Additionally, CSP captures atomic positions at a single instant in time, providing a snapshot of the structural state of the system. On the other hand, LENS analyzes the local system's dynamics by comparing changes in atomic neighborhoods between consecutive timeframes (i.e., LENS is a function of $\Delta$t), offering insight into how atomic environments evolve over time. 

To capture the overall system's response under strain, we calculated the total LENS (blue) and CSP (red) signals at each timestep by summing the individual signals from all atoms (Fig.~\ref{fig:fig2}b). At any given moment in the simulation, each atom has its own characteristic LENS and CSP values, reflecting its dynamic and structural state, respectively. By summing the LENS signals of all atoms at a specific timestep, we obtain the total LENS signal for the entire system at that moment. This summation is performed at every timestep, resulting in a time series of the total LENS signal throughout the simulation. The same method is applied to the CSP signals, allowing us to monitor both the dynamic (LENS) and structural (CSP) changes throughout the simulation.

Peaks in these cumulative LENS or CSP curves reflect concerted events, \textit{i.e.}, moments when many atoms exhibit simultaneous spikes in LENS or in the CSP signals. Furthermore, superimposition in the LENS and CSP peaks indicated events in time in which a large number of atoms underwent simultaneous structural (CSP) and dynamical (LENS) transitions, indicating large-scale structural changes. During the elastic phase ( $\sim$4 ns, \textit{cf.} Fig.~\ref{fig:fig1}b), neither LENS nor CSP show significant peaks, as local atomic movements emerged randomly and uncoordinated. LENS remains fairly stable, while CSP stays flat, reflecting preserved symmetry. At $\sim$4 ns, sharp peaks in both LENS and CSP signals mark the transition to plasticity, indicating simultaneous and coordinated atomic rearrangements. The coinciding spike in both descriptors indicates a collective structural event, where localized atomic reshuffling (captured by LENS) and broader symmetry-breaking (captured by CSP) occur simultaneously. Another set of peaks at $\sim$6 ns in Fig \ref{fig:fig2}b indicates a major dislocation event, characterized by collective atomic rearrangements and large-scale structural changes. After $\sim$6 ns, lower intensity LENS peaks indicate limited and localized movements,  while the fluctuating CSP signal reflects ongoing large-scale structural reconfigurations. Toward the end of the MD trajectory, the CSP fluctuations point to growing structural instability.

While the collective LENS and CSP values provide an overview of the system’s behaviour, analyzing their individual values reveals local atomic-level processes that remain hidden in the cumulative data. In contrast to the sum plots, the time-series of individual  $LENS_{i}$ and $CSP_{i}$ data (Fig.~\ref{fig:fig2}c and Fig.~\ref{fig:fig2}d, light blue and red data respectively) show how isolated atomic movements contribute to the system’s dynamic evolution. The accompanying KDE plots illustrate the distribution of descriptor values in the whole trajectory. Focusing first on LENS (Fig.~\ref{fig:fig2}c), we observe a ''forest'' of continuous, noisy individual peaks throughout the simulation, reflecting local dynamic atomic rearrangements. Such isolated movements are detected consistently but do not significantly alter the overall structure until many atoms move together in a coordinated way, causing the concerted spikes seen at $\sim$ 4 and $\sim$6 ns, that correspond to major dislocation events and to
the transition to the plastic phase. After these collective events, smaller fluctuations continue to be captured by the LENS signal, indicating ongoing localized atomic dynamics, even as the system stabilizes. CSP (Fig.~\ref{fig:fig2}d), in contrast, behaves differently. The individual $CSP_{i}$ signals remain low during the elastic phase but rise sharply during collective structural events, underpinning the plastic phase. Unlike $LENS_{i}$, which exhibit sharp short-lived peaks, $CSP_{i}$ values remain elevated after such events, indicating groups of atoms permanently shifted into asymmetric, defected configurations. This gradual increase in the CSP values during the plastic phase reflects the system's drift from its original crystalline symmetry. Major structural disruptions at $\sim$ 4 and $\sim$ 6 ns cause collective CSP peaks, but even after these events, CSP continues to grow, showing how the material’s overall symmetry progressively deteriorates as more and more atoms become structurally defected.

To explore how such local dynamic atomic rearrangements relate to structural disorder, we performed a correlation analysis between $LENS_{i}$ and $CSP_{i}$ data; Fig.~\ref{fig:fig2}e shows a scatter plot of the normalized $CSP_{i}$ and $LENS_{i}$ values throughout the simulation (note that $LENS_{i}(\Delta t)$ data are brought back by $(\Delta t )/2$ - $\Delta t$= 2 ps in the analysis- to match the $CSP_{i}$\cite{crippa2023machine}: see methods for details). The broad distribution of points reflects the complex, non-linear relationship between the local structural environment and their local dynamic as captured by the LENS e CSP descriptors during deformation.
The dashed lines at CSP = 0.4 (red) and LENS = 0.4 (blue) divide the plot into four distinct regions (where intrinsic vibrations/noise is clearly separated from relevant fluctuations) which correspond to different kinds of events. Here, we define as an ''event'' any atomic behavior that surpasses these thresholds, indicating significant local structural distortion (CSP $>$ 0.4) or statistically-relevant dynamic neighbor reshuffling fluctuations (LENS $>$ 0.4). In the upper-left quadrant (blue), two concentrations of points are identified by significant LENS activity without major structural distortion (low CSP)\cite{Crippa2023,crippa2023machine,LEAP,becchi2024layer}. The leftmost cluster in this quadrant is associated with atoms in the \textit{fcc} phase, where neighbor rearrangements occur with minimal structural distortion, while the rightmost more scattered cluster represents rearrangements in atoms within the \textit{hcp} phase of the stacking faults. The bottom-right quadrant (red) represents high CSP and low LENS values, indicating atoms that have lost their symmetry but exhibit little dynamic activity\cite{LEAP}. These atoms remain in disordered configurations without significant neighbor rearrangement, suggesting static structural disorder, \textit{i.e.}, persistent defects. Finally, the top-right quadrant (magenta) contains points where both CSP and LENS are high, identifying atoms undergoing both significant structural distortion and dynamic reshuffling. This quadrant highlights moments where both dynamic and structural changes are occurring at the same time, typically marking critical dominant events in the deformation process, as we will describe later.

Tracking the evolution of an individual atom through such 2D LENS-CSP space highlights how local dynamic and structural changes unfold during the metal deformation (Fig.~\ref{fig:fig2}f). We select an example of prototypical atoms that are part of a stacking fault formed during the dislocation event taking place at $\sim$6 ns (\textit{cf.} Fig.~\ref{fig:fig1}). For most of the simulation, this atom resides in the ''stable'' quadrant (bottom-left: \textit{i.e.}, both LENS and CSP values are below threshold), indicating minimal structural distortion and dynamic activity (intrinsic noise and vibrations). However, at $\sim$ 6 ns, the atom experiences a sharp change in both LENS and CSP signals, undergoing synchronized neighbor rearrangement and a sudden breaking of local structural symmetry, that corresponds to a dislocation event. This phase reflects the atom's abrupt drift from the ordered \textit{fcc} structure, with a high CSP value signaling significant structural disorder. After such a major event, the LENS value drops while the CSP signal remains high, as the atom transitions into the \textit{hcp} phase of the stacking fault, returning to a structurally stable but distorted state. Later, the atom returns to the \textit{fcc} phase with a minor LENS spike, suggesting that its local environment stabilizes without significant neighbor rearrangement, and it remains in this ordered state for the remainder of the simulation.

This specific journey is representative of general trends observed during dislocation events, as similar atom trajectories shown in the Supporting Information demonstrate. Critically, observing this trajectory in the 2D LENS-CSP space reveals correlations between local structural disorder and dynamic rearrangements that are not immediately apparent from the individual or global signals alone. The atomic-level resolution achieved in correlating CSP and LENS descriptors also allows for an analysis of their temporal relationship at the individual atom level (Fig.~\ref{fig:fig2}g). The histogram quantifies the time difference between the maximum values of CSP and LENS for each atom. By binning the data with a time-interval resolution of $\Delta t = 0.5 , \text{ns}$, we account for slight frame misalignments, ensuring that both correlated and non-correlated events are captured.
The domain identified by the black Gaussian includes all fluctuations from the LENS-CSP 2D plane (identified by purple point in panel e, where LENS and CSP have $\Delta$ t=0) along with system noise, which averages out at $\Delta t=0$. This domain represents the moments where dynamic rearrangements and structural disorder are synchronized, as well as the inherent noise in the system.
Importantly, the peak at approximately $\Delta \tilde{t} \sim 2$ ns, identified by the blue Gaussian, highlights a crucial observation: specific LENS fluctuations precede and lead to the formation of CSP defects. This characteristic delay ($\sim 2 ns$) between the precursor LENS fluctuation and the subsequent CSP defect formation suggests that atomic neighbor rearrangements detected by LENS are early indicators of structural disorder.  Further details on this mechanism will be discussed in the next section.

\subsection*{Local vs. collective defect dynamics in elastic deformation}

\begin{figure}[H]
\centering
\includegraphics[width=\linewidth]{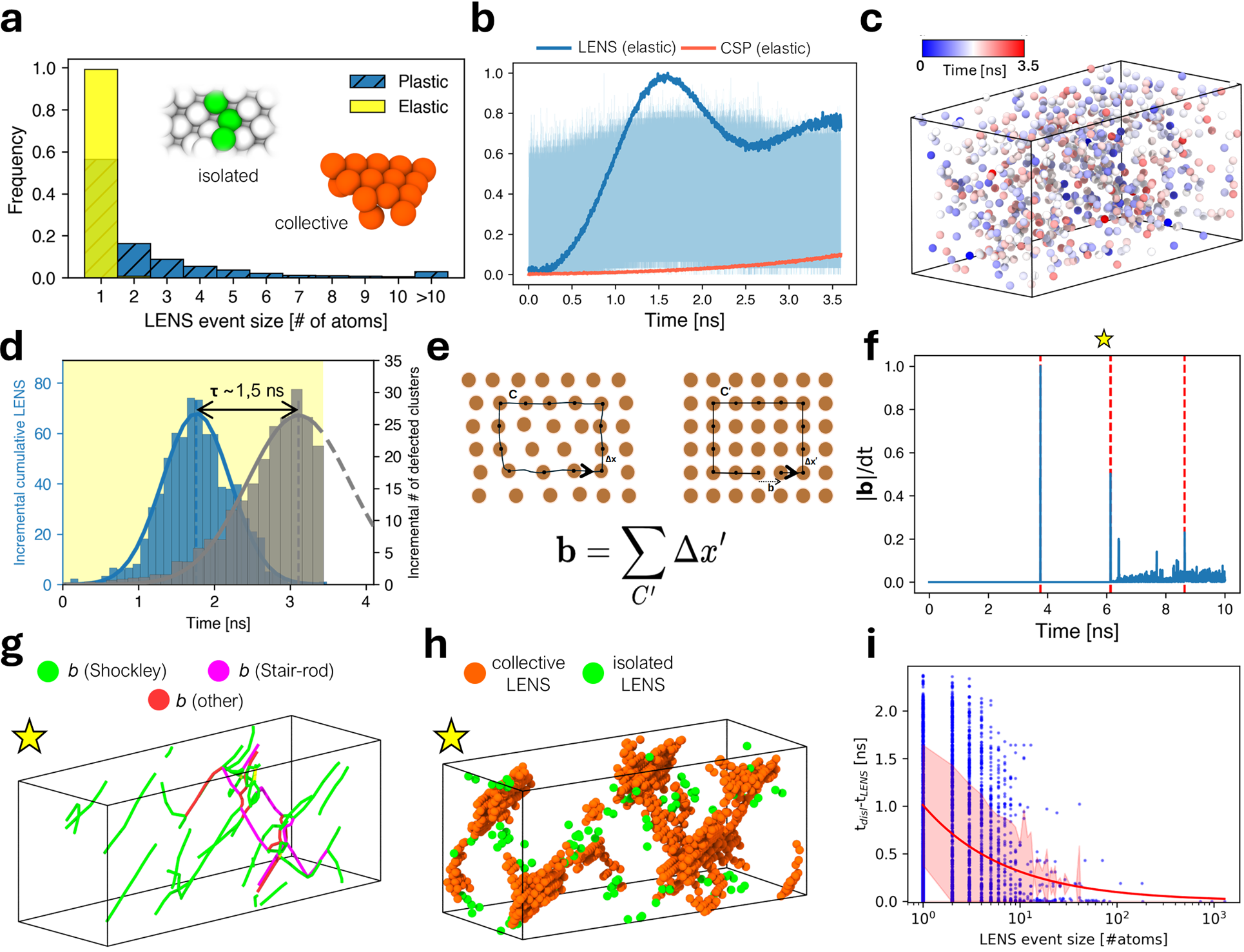}
\caption{Analysis of defect formation, dislocation behavior, and LENS events during tensile deformation of Cu.
(a) Distribution of LENS event sizes during the elastic (yellow) and plastic (blue) phases. The inset shows atomic configurations for isolated (green) and collective (orange) LENS events.
(b) Temporal evolution of LENS values (blue: summed over all atoms; cyan: individual atomic contributions) and CSP values (red: summed) during the elastic phase, showing numerous individual LENS peaks associated with rearrangement events.
(c) 3D visualization of the spatial distribution of LENS events during the elastic phase, color-coded by time.
(d) Cumulative LENS signal (blue) and cumulative number of nucleated defect clusters (gray) over time during the elastic phase.
(e) Schematic of the Burgers vector (\(\mathbf{b}\)), illustrating the lattice distortion due to dislocation loops, with the vector calculated by summing atomic displacements (\(\Delta \mathbf{x'}\)) around a closed loop. This vector characterizes the magnitude and direction of lattice distortion.
(f) Time derivative of the Burgers vector magnitude (\(\mathrm{d}|\mathbf{b}|/\mathrm{d}t\)), showing peaks corresponding to significant dislocation activity. The red dashed lines mark key dislocation activity events, while the yellow star indicates an exemplary point of interest.
(g) 3D visualization of Burgers vectors (\(\mathbf{b}\)) during an exemplary dislocation event, color coded according to the dislocation type.
(h) 3D visualization of LENS events during the same dislocation event of (g), color-coded to differentiate collective (orange) and isolated (green) events, demonstrating their spatial correlation with dislocations.
(i) Scatter plot depicting the relationship between LENS event size and the delay in dislocation formation (\(t_{\mathrm{b}} - t_{\mathrm{LENS}}\)), with larger LENS events showing no delay, indicating their role in driving immediate plastic activity. The red line highlights the general trend.
}
\label{fig:fig3}
\end{figure}

As a next step, we focus on interpreting the physical meaning of the dynamic phenomena captured by the LENS descriptor during the material's deformation. Specifically, we aim to differentiate between spatially localized LENS events, where individual atoms exhibit isolated signal spikes, and collective LENS events, where clusters of neighboring atoms (spatially correlated) spike simultaneously (temporally correlated), indicating wide dynamic activity. To achieve this, we identify atoms experiencing a statistically relevant LENS event at time $t_{i}$ and group them into clusters based on their proximity, \textit{i.e.}, atoms are considered part of the same cluster if their Euclidean distance is within a specified cutoff (2.8 \AA, corresponding to the first nearest neighbors shell). Given that the elastic phase exhibits more noise compared to the plastic phase, we applied a higher threshold of LENS $\ge$ 0.6 to better capture significant signals above the noise level. In contrast, for the plastic phase, which has a clearer signal, a lower threshold of LENS $\ge$ 0.4 was sufficient to differentiate meaningful events from the background noise. Fig.~\ref{fig:fig3}a shows the size distribution of the LENS event sizes observed along the simulations, with yellow bars representing the elastic phase and blue bars depicting the plastic phase. The decision to treat these two phases separately stems from their markedly different dynamic behaviors, as previously discussed (Fig.~\ref{fig:fig2}). While the elastic phase is characterized by isolated LENS atomic fluctuations, the plastic phase displays a broader distribution of LENS events, signaling a shift toward more collective, spatially correlated dynamic atomic events. This transition from isolated to collective events evidenced by LENS is a key distinction between the two phases, which deserves a detailed investigation

As said, the elastic phase is primarily characterized by isolated single-atom LENS fluctuations, with event sizes peaking sharply at 1 (Fig.~\ref{fig:fig3}a). As shown in Fig.~\ref{fig:fig3}b, the summed LENS signal in this phase remains relatively high despite the absence of any concerted peaks, as discussed previously. This is an indicator of localized atomic rearrangements, while the relatively flat CSP signal indicates the lattice structure remains overall stable. Despite the small, isolated nature of these fluctuations, they collectively contribute significantly to the cumulative LENS signal, highlighting the presence of local dynamic activity even in the absence of global structural reorganizations in the system. Throughout the elastic phase, LENS events show no spatial correlation, between local defects, or dynamic ''hot spots''\cite{Ngan2006}, continuously emerging and vanishing randomly across the lattice at different positions and times (Fig.~\ref{fig:fig3}c). This underscores the short-lived nature of such dynamic defects in this phase. These ''hot spots'' correspond to individual atomic fluctuations that do not directly lead to global structural changes, emphasizing that, although local dynamic activity is present at the atomic level, it does not compromise the overall integrity of the lattice.\cite{Ryu2011}

We aim to characterize the nature of these ephemeral defects from both dynamic and structural perspectives. In Figure~\ref{fig:fig3}d, we show the incremental cumulative LENS signal (blue) alongside the incremental number of detected defect clusters (grey bars) over time. The first major peak in the LENS signal marks a phase of intense localized atomic fluctuations preceding the formation of structural defects. The earlier rise in the LENS signal compared to the grey bars indicates a time delay between atomic rearrangements and defect nucleation.
Indeed, LENS fluctuations serve as early indicators of defect formation, signaling the onset of structural transformations. During the elastic phase, where defects accumulate in a spatially uncorrelated manner, these fluctuations allow us to track the evolution of structural changes. The most significant increase in the cumulative LENS signal occurs around $\sim$1.8 ns, highlighting a period of intense fluctuations. As the elastic phase progresses and local defects become more stable, the accumulation of LENS fluctuations gradually diminishes. This shift suggests that earlier fluctuations contributed to the formation of stable structural changes, marking an important phase in the material's transformation. Once defects stabilize, the LENS signal in those areas decreases, emphasizing its role in capturing transient atomic rearrangements that precede lasting structural modifications.
Moreover, LENS captures both the emergence and annihilation of defects, emphasizing the transient nature of local fluctuations. The increase in defect clusters doesn't imply that all previously formed defects persist; instead, the system experiences continuous cycles of defect formation and annihilation, resulting in a net increase in defect clusters over time.
As we move into the plastic phase, random defect accumulation gives way to dislocation formation. Despite the decline in the LENS signal, indicating a reduction in atomic rearrangements, defect nucleation continues, as indicated by the steady rise of the grey bars. This suggests that structural changes persist even though dynamic fluctuations decrease. The maximum increment in defect formation occurs around $\sim$3 ns, marking the transition from elastic to plastic deformation. Interestingly, the delay between the peak of the cumulative LENS increment (blu) and the maximum in defect formation (grey) is approximately $\tau \sim 1.5$ ns. This delay is consistent with our previous findings in Figure~\ref{fig:fig2}g, where we observed a similar time lag between the dynamic fluctuations detected by LENS and the onset of structural disorder. This highlights the role of LENS fluctuations as early indicators of structural transformations, with persistent defect nucleation occurring after these dynamic events. This progression from transient atomic rearrangements to stable defects emphasizes how localized dynamics drive the material toward plasticity, providing a detailed understanding of the processes preceding large-scale structural transformations.
We note that the time delay observed between transient fluctuations detected by LENS and the onset of plasticity is influenced by the strain rate applied in the simulations, with these timescales expected to scale approximately linearly with the strain rate.
As a next step, we discuss the result obtained for the plastic phase of the simulation. As shown in Fig.~\ref{fig:fig3}a, the distribution of LENS event sizes in this phase (blue bars) still sees considerable presence of events with the mode of 1, indicating the continued presence of isolated, individual dynamic fluctuations similar to those observed in the elastic phase (that anticipate and trigger lager-size defect). However, a key difference is the emergence of larger events, with the final column (size $\ge$ 10 atoms) representing collective LENS events involving up to hundreds of atoms. These larger-scale events highlight the transition to the plastic phase. Indeed, structural changes in the lattice leading to plastic deformation are driven by the formation and propagation of dislocations, which are collective line defects that move along specific crystallographic planes to allow slip and relieve stress. Dislocations introduce irregularities in the lattice that may be efficiently quantified by, \textit{e.g.}, the Burgers vector ($\mathbf{b}$), which defines both the magnitude and direction of the lattice distortion caused by the dislocation (Fig.~\ref{fig:fig3}e). To track and quantify these distortions over time, we employ the DXA algorithm\cite{Stukowski2012}, an established method for identifying and measuring the Burgers vector in dislocated systems. In this study, we extend this approach by analyzing the time derivative of the Burgers vector, as shown in Fig.~\ref{fig:fig3}f. This derivative form, \(\frac{d|\mathbf{b}|}{dt}\), provides a time-resolved measure of dislocation activity, enabling us to identify the precise moments in which the dislocations nucleate or propagate within the system. The resulting time series reveals distinct peaks that correspond to dislocation events, giving us a direct way to monitor the progression of structural changes during plastic flow. We then identified and recorded the time instants at which significant variations in the Burgers vector occurred, marked by red dashed lines in the plot (see Fig.\ref{fig:fig3}F). In this graph, the depicted quantity represents the sum of the Burgers vectors of all dislocation lines within the simulation cell. As expected, these plastic events are simultaneous with the stress and potential energy drops illustrated in Fig.~\ref{fig:fig1}b.

Concerning the plastic phase, our goal is to correlate atomic-scale dynamic fluctuations, captured by LENS, with system-wide collective events, such as dislocation emergence and activity. In particular, while LENS detects local dynamic fluctuations at the atomic level, the Burgers vector describes larger-scale lattice distortions. By correlating peaks in the time derivative of the Burgers vector to variations in the LENS signal, we can connect local atomic dynamics and defect activity with the onset of large-scale collective structural rearrangements. We begin by exploring this correlation qualitatively. Fig.~\ref{fig:fig3}g shows the spatial distribution of dislocations around 6 ns, with various types identified using the DXA algorithm,\cite{Stukowski2012} including Shockley (green), Stair-rod (magenta), and other dislocation types (red).\cite{stukowski2012structure} Simultaneously, as previously described, the journey of an individual atom on the 2D LENS-CSP space in Fig.~\ref{fig:fig2}f reveals a rise in both descriptors at $\sim$ 6 ns,  precisely corresponding to the second dislocation event.

The snapshot in Fig.~\ref{fig:fig3}h further supports this correlation; it shows LENS activity at the same frame time of Fig. ~\ref{fig:fig3}g, showing a clear spatial correlation between Burgers vectors and atoms undergoing LENS events. Notably, the majority of these atoms belong to a single large-scale LENS cluster, indicating that this dislocation event involves significant collective atomic rearrangements. This correlation between the dislocation network and the LENS cluster underlines how large-scale dislocations are phenomena characterized by collective atomic dynamics. To attain a more quantitative connection in  Fig.~\ref{fig:fig3}i we estimated the delay time between collective dislocation events (as marked by the red dashed lines in Fig.~\ref{fig:fig3}f, and the LENS fluctuations they originate from (plotted as a function of the event's size). The data show how, for large-scale events involving hundreds-to-thousands of atoms, the delay time tends to $\sim 0$, indicating that a dislocation is captured exactly as a collective lens event.  In contrast, smaller (more local) LENS events, particularly those involving only a few atoms, exhibit much greater variability in the delay time, ranging from 0 to 2.5 ns. This indicates how localized LENS atomic fluctuations (and the local defects that these may create) are not as tightly correlated with dislocations, underscoring the importance of collective dynamics in triggering the plastic phase and driving plastic flow.
Furthermore, this provides a general difference between the elastic and the plastic phases. In the elastic phase, the local LENS fluctuations and the formation of local structural defects is temporarily misaligned, with LENS fluctuation anticipating the formation of local defects that become dynamically persistent after $\sim$1.5 ns (see Fig.\ref{fig:fig3}d). When such local defects become so numerous that their spatial correlation is unavoidable, they trigger the formation of dislocations, that is characterized by the fact that LENS fluctuation exactly corresponds to dynamic evolution of such structural defects (the lag time between collective LENS fluctuations and the dislocation movements is equal to 0).
Noteworthy, such a distinction between the structural and dynamical features characterizing the transition from the elastic to the plastic phase is obtained here by using a single descriptor and its abstract basic concepts, specifically the detailed analysis of LENS local fluctuations and their correlation in space and time.
The distinction between local and collective dynamical events will be further explored in the next section, where we investigate how such complex and diverse atomic-level dynamics characterize structural failure during fracture.


\subsection*{Atomic dynamics during and post fracture}

Building on our analysis of local and collective dynamics during plastic deformation, we now turn to the investigation of how these dynamics evolve during metal structural failure. Fracture, as a final stage of deformation, represents a critical transition where accumulated atomic disordering culminates in the formation of new surfaces and defects. The insights gained from dynamic descriptors are particularly valuable in this context, as they allow us to probe not only the onset of fracture but also the dynamic processes that persist beyond it. To explore these phenomena, we conducted a separate set of simulations under NVT conditions where a constant strain rate is applied along the \textit{x} direction (\textit{cf.} Methods for details), using 6 independent replicas as before. Aggregated statistics are presented, with individual plots provided in the Supplementary Information. As shown in Fig.~\ref{fig:fig4}, the imposed strain leads to the nucleation of defects and ultimately the creation of fracture surfaces. These new surfaces raise the system's energy, driving the material into a metastable state where localized atomic rearrangements continue as the system seeks to minimize its energy. Fig.~\ref{fig:fig4}a depicts this transition, with the material evolving from an intact structure to a fractured state. Correspondingly, the stress-strain curve in Fig.~\ref{fig:fig4}b highlights the sharp drop in stress at the fracture point, followed by a modest elastic recovery and subsequent stabilization. This partial recovery reflects the inability of the material to return to its pre-fractured energy state, as the formation of defects and new surfaces has irreversibly altered its structure and energetics.

We applied the same analysis used in the previous sections to analyze the MD trajectories obtained from these fractured simulations. Fig.~\ref{fig:fig4}c shows the cumulative LENS (blue) and CSP (red) signals (sum of the signals of all atoms) over time. As described earlier, peaks in these plots reflect concerted events, indicating large-scale structural changes. Up to $\sim$4 ns, the behavior of the cumulative LENS and CSP signals mirrors the previous observations for the elastic phase (see Fig.~\ref{fig:fig2}b). This similarity is expected, as up to $\sim$4.5 ns the system is still within the elastic regime(see Fig.~\ref{fig:fig4}b). Fig.~\ref{fig:fig4}d also reflects this behavior, with individual LENS signals following the same pattern of isolated local atomic rearrangements before the elastic-plastic transition as described previously. At $\sim$ 4.5 ns, both Fig.~\ref{fig:fig4}c and Fig.~\ref{fig:fig4}d display a peak in the LENS signal corresponding to the emergence of a dislocation event. This peak is accompanied by the characteristic stress and potential energy drop shown in Fig.~\ref{fig:fig4}b, further confirming the onset of plastic deformation. Shortly after, at 5.5 ns, the system undergoes a major structural transformation as the fracture event occurs, which is marked by an intense spike in the LENS signal, visible in both Fig.~\ref{fig:fig4}c and Fig.~\ref{fig:fig4}d. This LENS peak is significantly larger than the dislocation-related peaks, emphasizing the magnitude of the atomic rearrangements during fracture. In contrast, the CSP signal exhibits only a modest increase, indicating that global symmetry changes (structural rearrangements) are less abrupt. This difference in magnitude and rate between the LENS and CSP values leading up to the fracture indicates that local atomic reshuffling, rather than broader and persistent structural changes, are the primary drivers of this process. Fig.~\ref{fig:fig4}e shows MD snapshots where 
the atoms are colored based on the LENS fluctuations involving them. These snapshots allow visualizing the spatial distribution of LENS fluctuations at the moment of fracture.

Brighter colors in the image highlight concerted dynamic activity concentrated along the fracture surfaces, where defects and irregularities formed due to the ductile fracture exhibit significantly higher LENS values and increased atomic mobility. Altogether, Fig.~\ref{fig:fig4}c, Fig.~\ref{fig:fig4}d, and Fig.~\ref{fig:fig4}e demonstrate that the fracture process involves a highly focused concerted increase in atomic mobility, as the material transitions through the fracture event. 
The post-fracture behavior is also particularly interesting from the dynamics point of view. As seen in Fig.~\ref{fig:fig4}c, both LENS and CSP values stabilize, but the LENS signal remains higher, indicating that the residual local atomic activity is present along the newly formed fracture surfaces. CSP, being system-specific and sensitive primarily to defect-driven disorder, offers little insight after fracture as the bulk material returns to a stable, defect-free \textit{fcc} crystal following elastic recovery. To better understand this persistence of localized activity post-fracture, Fig.~\ref{fig:fig4}d continues to show the time evolution of individual LENS values. While the aggregated signals level off after fracture, atomic rearrangements persist at the fracture surfaces, particularly at locations where reduced coordination leads to increased atomic mobility. These localized fluctuations suggest that the system has not fully relaxed at the atomic scale, and atoms continue to rearrange along the fracture interfaces to minimize the system's energy. This persistence of atomic activity is further confirmed by the large energy fluctuations observed in Fig.~\ref{fig:fig4}b.

This discrepancy underscores the potential of analysis based on abstract descriptors such as LENS, to capture a broad range of local and collective dynamic processes, providing a unified picture of metal under stress, from the elastic phase to post-fracture surface dynamics.
Notably, this behavior is similar to what has been recently observed in metal surfaces, where local atomic diffusion and sliding can occur even at relatively low temperatures. These are captured by LENS signals but not by structural descriptors\cite{LEAP}.
This is reasonable in this case, as when the metal breaks, two new surfaces are formed, and their dynamics are more similar to those of typical metal surfaces rather than the relatively static dynamics characteristic of the metal bulk. In this specific scenario, such residual dynamics on the surface are facilitated even at 300 K since we are observing the system immediately after the surface breakage (while, over a longer time scale, the surface is expected to reach a more stable configuration).

Upon breakage, the two new surfaces undergo elastic relaxation, and even at low temperatures, on the scale of nanoseconds, this results in the typical signatures of surface dynamics — such as local atomic sliding and diffusion—captured by LENS fluctuations, but not by structural descriptors. These behaviors are characteristic of metal dynamics typically observed at higher temperatures.\cite{cioni2023innate,Crippa2023, Caruso2023}
In a similar manner to our previous analysis, Fig.~\ref{fig:fig4}f reports the distribution of LENS event sizes before and after the fracture, revealing important shifts in the underlying atomic dynamics. Using the same approach as in Fig.~\ref{fig:fig3}a, we renormalized the LENS signals separately for the pre-fracture and post-fracture phases, applying different thresholds for event detection: LENS $>$ 0.4 for the pre-fracture phase and 0.6 for the post-fracture phase. Before the fracture, the system is dominated by larger atomic clusters, with groups of atoms moving collectively as the material undergoes stress. This reflects the coordinated atomic movements over extended regions, a hallmark of the material's response during pre-fracture plastic deformation. In this phase, as the metal begins to separate into distinct sections, larger LENS events capture the increasing spatial coordination of atomic rearrangements, driven primarily by the propagation and interaction of dislocations.
Following the fracture, we observe a striking transition, and the LENS distribution shifts toward much smaller event sizes, typically involving just $\sim$ 1-5 atoms. This shift again signifies the presence of the fundamental transition in the nature of atomic dynamics explained above. Instead of coordinated movements across large domains, the material’s response becomes highly localized, confined to surface-level rearrangements. This observation is critical, as it reveals that the material has lost its bulk cohesion, and dynamic activity is now concentrated along the newly formed fracture that, immediately after rupture, are defected and are underpinning elastic relax and shows (even at 300 K) characteristic signatures of the local atomic dynamics and diffusions that are typical of these metal surfaces at higher termperatures.\cite{LEAP,becchi2024layer,crippa2023machine,Crippa2023}
Finally, Fig.~\ref{fig:fig4}g provides a direct view of the system after the fracture event, corresponding to step 2 in the stress-strain curve. We focus on a prototypical example that exhibits a dynamic behavior diffusing on the exposed surface after rupture (trajectory in blue in the inset). Such dynamical surface, marked by high LENS values, have reduced coordination and enhanced mobility. Despite the relatively low temperature of 300 K, in fact, these atomic rearrangements persist as the surface is decreasing its energy and the atoms adjust to the new configuration (elastic relax). The low temperature (300 K) in these simulations slows down the relaxation process, and the defected surfaces require a longer time than that accessible by these simulations to reach a new stable state. The LENS-based color map further distinguishes the highly active fractured region from the bulk-like, stable regions, which exhibit low LENS values and remain largely static. This observation reinforces the notion that the surface remains dynamically active, even post-fracture, as isolated atoms continue to search for energetically favorable configurations. The dynamic behavior observed on the surface mirrors similar findings in metal nanoparticles and \textit{fcc}-Cu surfaces\cite{cioni2023innate,rapetti2023machine,cioni2024sampling,crippa2023machine, Crippa2023}, where surface atoms maintain mobility well below the material's melting point. All in all, this suggests that the microscopic dynamic behavior plays a critical role in how material behaves following to fracture, with the defects acting as a driving force for sustained atomic rearrangements.
It is important to remark that the fracture phenomenon observed in the simulation does not reflect realistic material behavior, as it is a direct consequence of the computational approach tailored to induce fracture quickly. However, the objective of this simulation was not to analyze the detailed dynamics of fracture propagation but rather to investigate the atomic dynamics of the newly formed fractured surfaces during the elastic return phase. These surface dynamics, often overlooked in traditional studies, provide valuable insights into the stabilization and rearrangement mechanisms that follow fracture.
All these data confirm that there is an intricate relationship between the degree of ''collectivity'' in defect dynamics, \textit{i.e.}, the size of the LENS event, and the different phases (elastic, plastic, fracture) of the metal's behavior under deformation. The transitions from collective and back to local behaviors after rupture are therefore key aspects in  understanding how metals respond to mechanical stress. They underscore the dual nature of atomic dynamics: collective during the deformation phase, where large-scale movements dominate, and localized when the material is not in plastic phase and dislocation are quiescent (\textit{i.e.}, in the preliminary elastic phase and after fracture). These results highlight how LENS can effectively capture the evolving emergence and amplification of defect dynamics across all phases of deformation, providing insight into both bulk and surface-level atomic rearrangements without any necessary prior assumption: \textit{i.g.}, it allows to distinguish elastic from plastic phases due to the different spatial/temporal correlations between LENS fluctuation, and without the need to use \textit{ad hoc} descriptors/approaches that imply a prior knowledge of the fact that such two two different phases are effectively present (\textit{e.g.}, CSP and Burgers vector). Experimental validation and the integration of experimental data to support this study remain highly challenging. Nevertheless, existing evidence\cite{Barr2023,cioni2024sampling} corroborated by experimental observations demonstrates that spontaneous nanoscale dynamics, even at low temperatures, can significantly influence material behavior and properties.

\begin{figure}[H]
\centering
\includegraphics[width=0.91\linewidth]{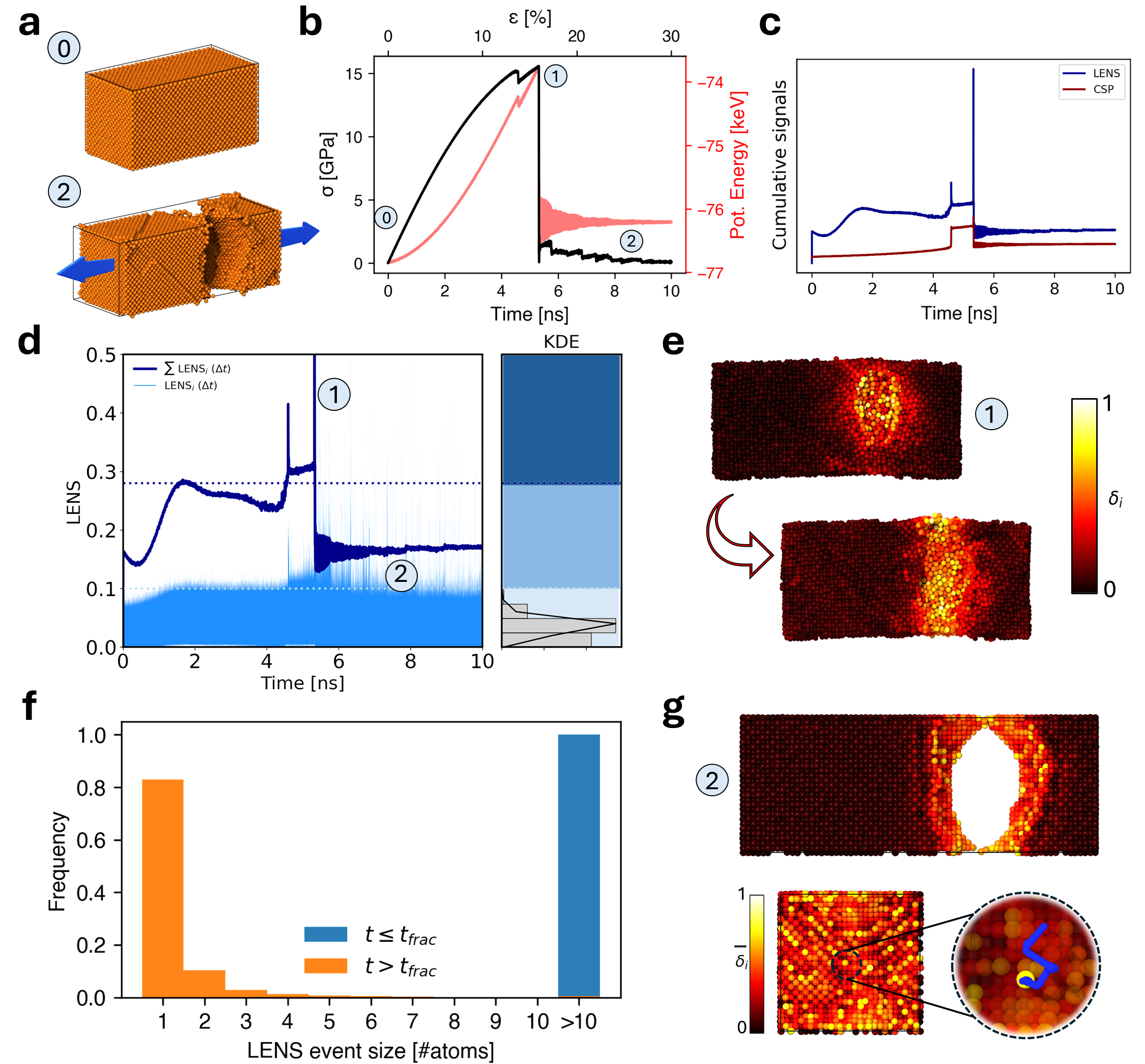}
\caption{
Atomic rearrangement and fracture behavior during catastrophic failure of Cu.
(a) Initial (0) and fractured (2) configurations of the Cu system under tensile loading.
(b) Stress-strain curve (black) and potential energy evolution (red) over time, illustrating elastic, plastic, and post-fracture regimes. Labels (1) and (2) correspond to the configurations shown in (a), (e), and (g).
(c) Temporal evolution of the summed LENS (blue) and CSP (red) values during deformation, with a sharp LENS peak marking the fracture event.
(d) Time evolution of individual atomic LENS values (\(\delta_i\)), with the Kernel Density Estimate (KDE) plot on the right representing the distribution of LENS values over time. The summed curve from (c) is overlaid for reference. Peaks labeled (1) and (2) indicate significant atomic rearrangements that match the stages identified in (b). 
(e) Visualization of atomic rearrangements at two time points for the event labeled as (1): before fracture (top) and during fracture (bottom). Atoms are colored by their LENS value (\(\delta_i\)), highlighting regions of heightened atomic activity.
(f) Distribution of LENS event sizes before fracture (orange) and after fracture (blue), showing the prevalence of individual events in the post-fracture regime.
(g) Visualization of the fractured system after deformation, including a zoomed-in view showing atomic motion along the fracture surface, represented by blue trajectories. The color scale indicates the average LENS values (\(\bar{\delta}\)) in the post-fracture phase.
}

\label{fig:fig4}
\end{figure}

\section*{Conclusions}
In this work, we combined dynamic and structural descriptors, \textit{i.e.}, LENS, CPS and Burgers vector to analyze atomistic molecular dynamics simulations and to investigate fundamental microscopic mechanisms driving plastic deformation and dislocation nucleation in copper under uniaxial tensile stress. In particular, by leveraging the agnostic LENS descriptor of local atomic dynamics, we tracked both localized and collective atomic rearrangements throughout the material’s elastic, plastic, and post-fracture phases, providing microscopic-level predictive insights into key steps and phenomena controlling dislocation formation and fracture. Despite building solely on abstract concepts, \textit{e.g.}, those of local dynamical fluctuations and of correlations of local fluctuations in space and time, this approach enables a comprehensive understanding of how dynamic neighbor rearrangements evolve from isolated (non-correlated) atomic fluctuations into large-scale collective structural transformations.

During the elastic phase, LENS reveals the emergence and multiplication of ephemeral localized defects that are continuously formed and annihilated in the system. As these defects become more and more dynamically persistent (their LENS signal stabilizes), and the system approaches its elastic limit, where such reversible, ephemeral atomic fluctuations transition into more frequent defect nucleation and growth. This process marks the system’s approach to plastic flow, providing a dynamic view where (LENS) local dynamical fluctuations are the precursors of permanent structural transformations. 

As the metal transitions into the plastic phase, we observe a clear shift from localized atomic fluctuations to collective defect dynamics, where clusters of atoms move in concert. This collective behavior is strongly correlated with dislocation nucleation, underscoring that the scale and intensity of these rearrangements are key drivers of plasticity. This is captured as a transition in the spatiotemporal correlations between local LENS atomic fluctuations that, in the plastic phase, appear simultaneously in time and correlated in space (planes sliding). While LENS tracks the dynamic rearrangements that anticipate local defect formation and accompany defect propagation, CSP captures the progressive breakdown of lattice symmetry (appearance of persistent structural defects) as the material enters in the plastic phase.

Post-fracture analysis further reveals that localized atomic activity persists along newly formed surfaces, which are subject to elastic relaxation. The continuous atomic dynamic rearrangements that are observed in these nanosecond-scale simulations immediately after fracture are consistent with, \textit{e.g.}, the atomic mobility observed in various types of (Cu) metal surfaces below melting at much higher (close to the Huttig) temperature.\cite{cioni2023innate,crippa2023machine,Crippa2023,rapetti2023machine} Despite here we are at a much lower temperature, such atomic dynamics is driven by defects that, immediately after the ductile fracture, are present in the newly formed surface (making them higher in energy than hypothetically perfect Cu surfaces at 300 K), and that act as seeds for atomic diffusion, sliding, and rearrangements. While the surfaces reconstruct and stabilize over time, this evidence highlights the prominent key role of atomic-scale dynamic processes in post-fracture material recovery.

The analyses conducted through these three phases have enabled the characterization of key dynamic phenomena, including: (i) the early-stage events leading to defect nucleation, (ii) the propagation of collective defects, and (iii) the residual dynamics following fracture. This has been made possible by complementing the LENS descriptor with traditional analysis methods such as CSP, allowing for the detection of fluctuations that precede the formation of stable defects. Studying these local fluctuations before the emergence of persistent defects could provide a foundation for identifying causal links in these systems.\cite{LEAP,DelTatto2024}

From a broader perspective, this framework offers new opportunities to understand the often complex interplay existing between local and collective dynamical events, and to compare it thanks to the abstract character of the analysis, between different metals and in systems of completely different nature \textit{e.g.}, polycrystalline systems, alloys, systems under extreme conditions, etc. Future studies will further explore its applicability to more complex material systems. Additionally, integrating AI-based techniques, such as convolutional neural networks (CNNs), could enhance the automated detection and classification of atomic-scale defects, further expanding the analytical capabilities of this approach.

\section{Methods}
\subsection*{MD simulation}
All simulations to investigate the mechanical deformation of a Copper system were conducted using the LAMMPS\cite{LAMMPS} software (version 2 August 2023). The simulated metal contained 21,952 Cu atoms, generated by replicating a face-centered cubic \textit{fcc} unit cell of Copper, respectively 28, 14, and 14 times along the \textit{x}, \textit{y}, and \textit{z} directions, to create the full simulation box. The simulation cell was simulated with periodic boundary conditions in all three dimensions, in such a way to effectively model the behavior of a portion of the bulk of a Cu crystal under tensile stress and deformation.

The interatomic interactions were modeled using the embedded atom method (EAM) with the Mishin copper potential.\cite{Mishin2001} Before starting the deformation simulations, the system was energy minimized using the conjugate gradient method. The minimization was performed with a stopping criterion of \(1.0 \times 10^{-15}\) energy tolerance, \(1.0 \times 10^{-15}\) force tolerance, and a maximum of 1000 iterations. The system is then equilibrated at 300 K and 1 bar for 1 ns. Temperature and pressure are kept constant using the Nosé-Hoover thermostat and barostat.\cite{Shinoda2004,Martyna1994} For each replica, initial velocities were drawn randomly from a Maxwell-Boltzmann distribution. Following the equilibration, two different protocols were employed for the deformation simulations, which are detailed below.

\textbf{NPT deformation simulations} Uniaxial strain was applied along the x-axis at a strain rate of \(3 \times 10^{7}\)$s^{-1}$, using the \texttt{fix deform} command. Stress and strain data were collected every 2 ps throughout the simulation, which ran for 10 nanoseconds. The other two dimensions were allowed to adjust dynamically to the imposed strain using an anisotropic Nosé-Hoover barostat, with zero pressure applied in the two directions orthogonal to the strain. Temperature was maintained constant at 300 K using the Nosé-Hoover thermostat. 

\textbf{NVT deformation simulations} As in the previous setup, uniaxial strain was applied along the x-axis at a strain rate of \(3 \times 10^{7}\)$s^{-1}$, using the \texttt{fix deform} command. Stress and strain data were collected throughout the simulation, which ran for 10 nanoseconds. In this protocol, orthogonal dimensions are kept fixed using the \texttt{fix nvt/sllod} command. Temperature was maintained at 300 K using the Nosé-Hoover thermostat with the SLLOD implementation.\cite{Todd2017}

\textbf{LENS analyses} For each atom \textit{i} in the system, the LENS value (fluctuation) is computed at each sampling time-interval ($\Delta t = 2 ps$ in all analyses reported herein) over time along the entire trajectory. Let $C^{t}_{i}$ represent an array containing the identities (IDs) of atoms surrounding the center \textit{i} within a sphere of radius $r_{cut}$ at the time step $t$. The LENS value at any time $t +\Delta{t}$, denoted as $\delta_{i}$, is defined as: 
\begin{equation}
\centering
    \delta_i^{t+\Delta t}=\frac{\#(C_i^{t} \bigcup C_i^{t+\Delta t} - C_i^{t} \bigcap C_i^{t+\Delta t})}{\#(C_i^t+C_i^{t+\Delta t})},
\label{eq:METHODLENS}
\end{equation}
where the union and intersection in the numerator represent the neighbor IDs within $r_{cut}$ from atom \textit{i} at times $t$ and $t + \Delta{t}$, respectively. This formulation allows $\delta_{i}(t)$ to capture changes in the local environment of atom \textit{i}, with values ranging from 0 for stable neighborhoods to 1 for the maximum dynamical ones. In our analysis, the $r_{cut}$ within which the atomic dynamics is monitored is equal to 8 $\AA$ (corresponding to the $\sim \mathrm{3^{rd}}$ atom neighbor distance). We conducted additional tests using smaller cutoffs, such as the first nearest-neighbor distance (3.1 $\AA$), as well as larger values. These tests consistently confirmed the observed trends across all cases. The choice of 8 $\AA$ represents a compromise between computational efficiency and the need to probe long-range spatial correlations.

\section*{Data availability}

Complete data and materials pertaining to the atomistic simulations and data analyses conducted herein (input files, model files, raw data, analysis tools, etc.) are available at  \url{https://github.com/GMPavanLab/DefectsTracker_METALS} (this link will be replaced with a definitive Zenodo link upon acceptance of the final version of this paper). Other information needed is available from the corresponding author upon reasonable request.

\bibliography{biblio}

\section*{Acknowledgements}

G.M.P. acknowledges the funding received by the ERC under the European Union’s Horizon 2020 research and innovation program (grant agreement no. 818776 - DYNAPOL). The authors also acknowledge the computational resources provided by CINECA and by HPC@POLITO (http://www.hpc.polito.it).

\section*{Author contributions}

G.M.P. conceived this research. M.P., M.C. and M.D.P. developed the molecular models. M.P. and M.C. performed the simulations. M.P., M.C., M.D.P. and G.M.P. analysed the results. M.D.P. and G.M.P. supervised the work. M.P., M.C., M.D.P. and G.M.P. wrote the manuscript.

\section*{Competing interests}

Authors declare no competing interests.

\end{document}


\maketitle

\newpage

\begin{figure}[H]
\centering\includegraphics[trim={0cm 0cm 0cm 0cm},width=\textwidth]{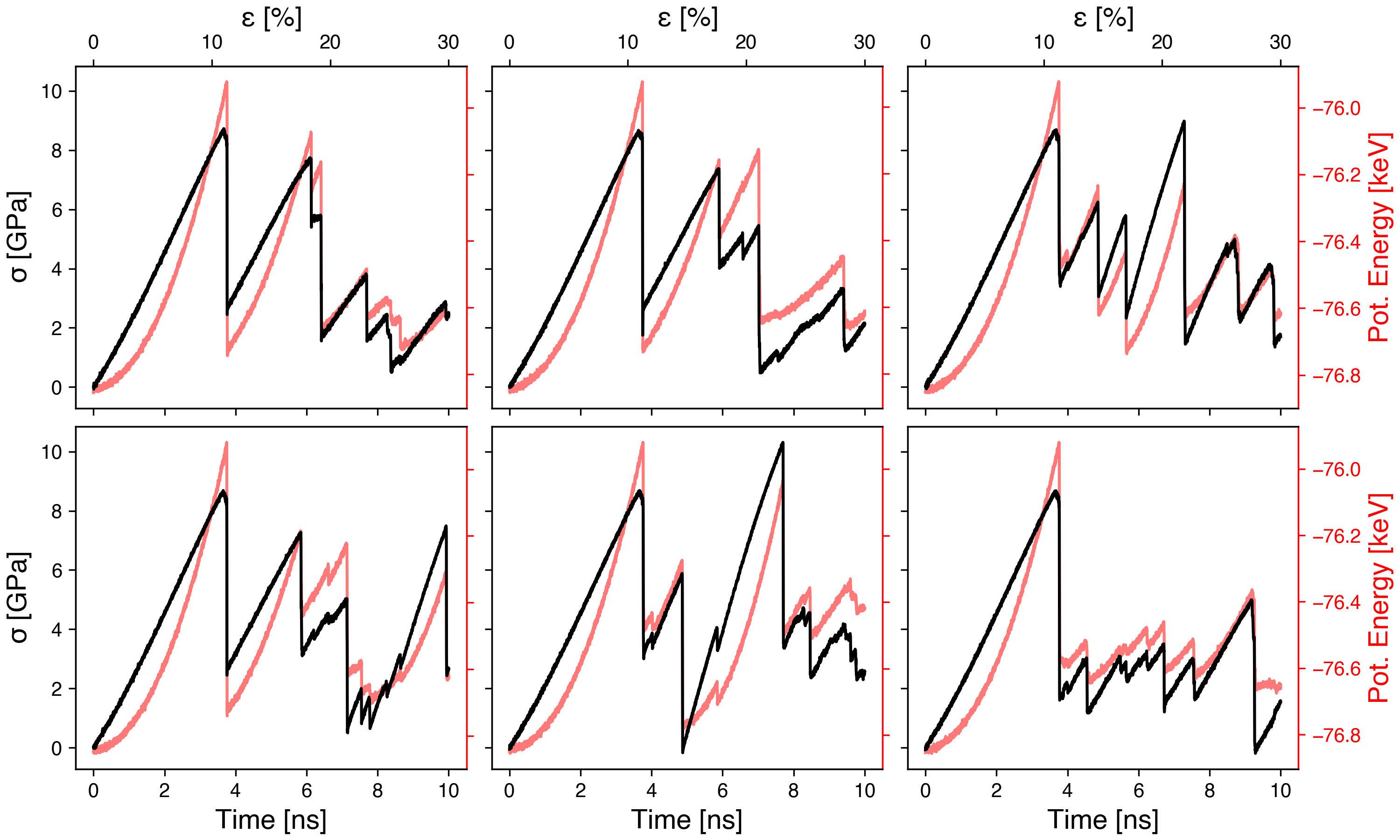}
\caption{Plots of the stress (black curves) and potential energy (red curves) as a function of time for the 6 replica simulations of the system in NPT conditions with constant applied strain rate.}
\label{fig:SSNPT}
\end{figure}
 \begin{figure}[H]

\centering\includegraphics[trim={0cm 0cm 0cm 0cm},width=\textwidth]{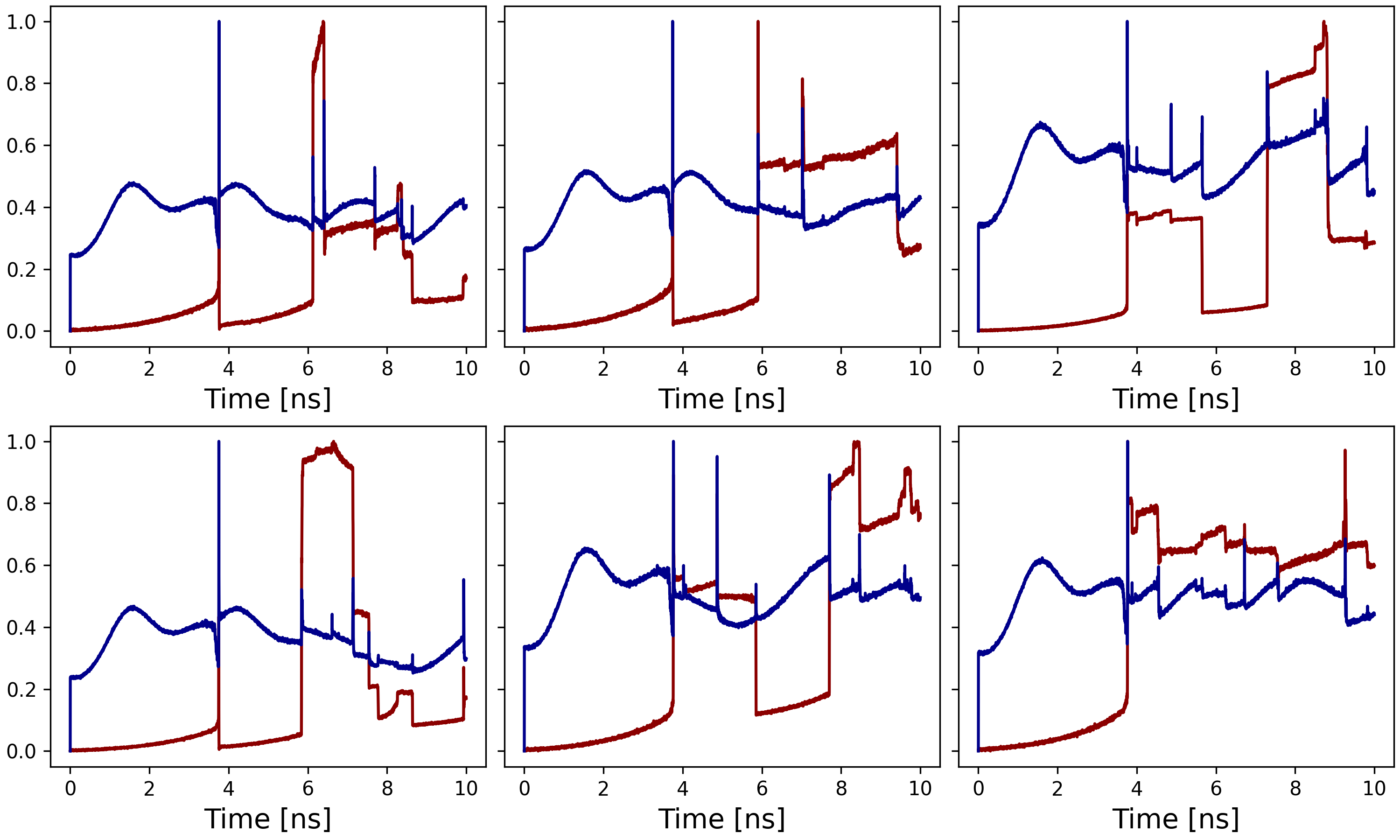}
\caption{Plots of the normalized summed LENS signal (blue lines) and summed CSP signal (red cueves) as a function of time for the 6 replica simulations of the system in NPT conditions with constant applied strain rate.}
\label{fig:SUMNPT}
\end{figure} 

\begin{figure}[H]
\centering\includegraphics[trim={0cm 0cm 0cm 0cm},width=\textwidth]{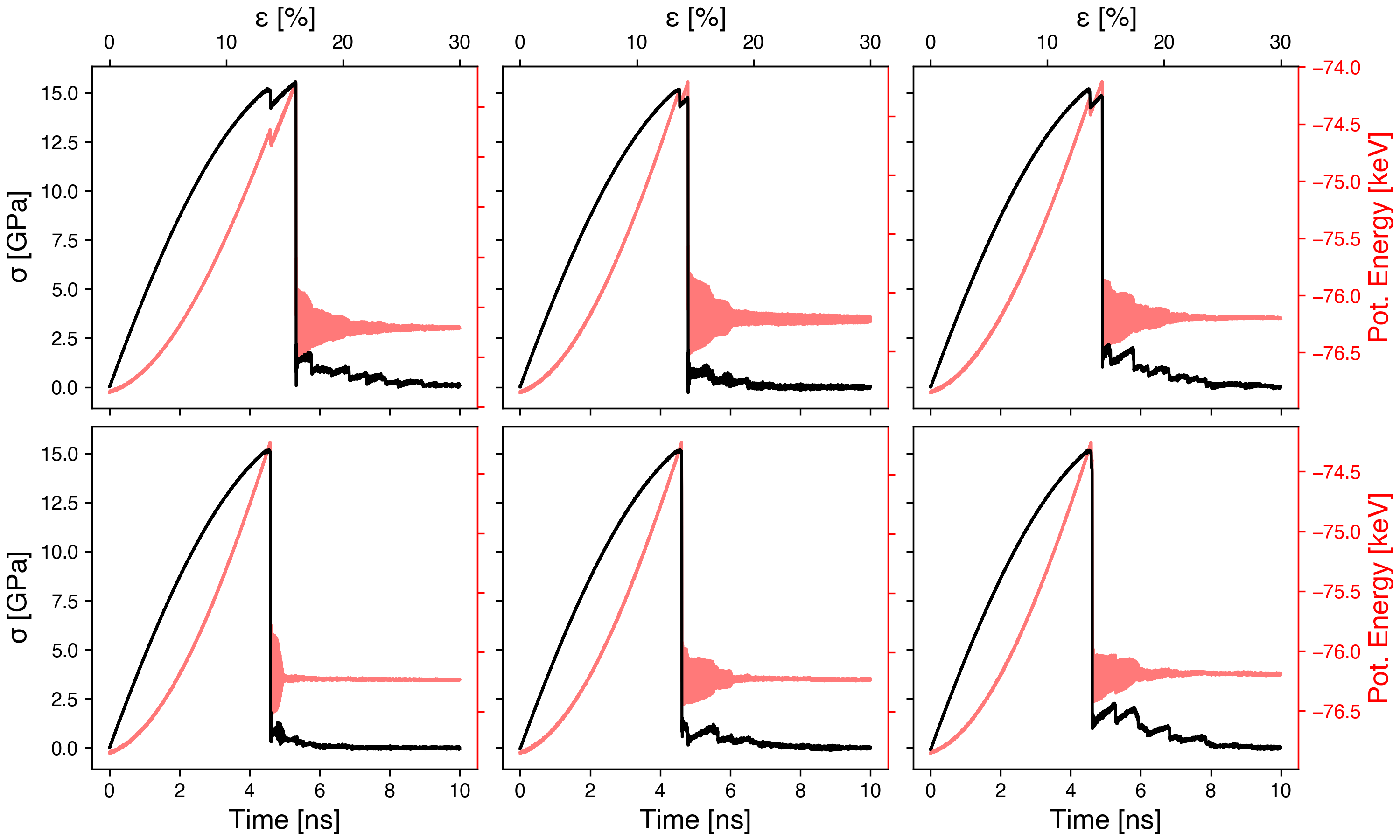}
\caption{Plots of the stress (black curves) and potential energy (red curves) as a function of time for the 6 replica simulations of the system in NVT conditions with constant applied strain rate.}
\label{fig:SSNVT}
\end{figure}

\begin{figure}[H]
\centering\includegraphics[trim={0cm 0cm 0cm 0cm},width=\textwidth]{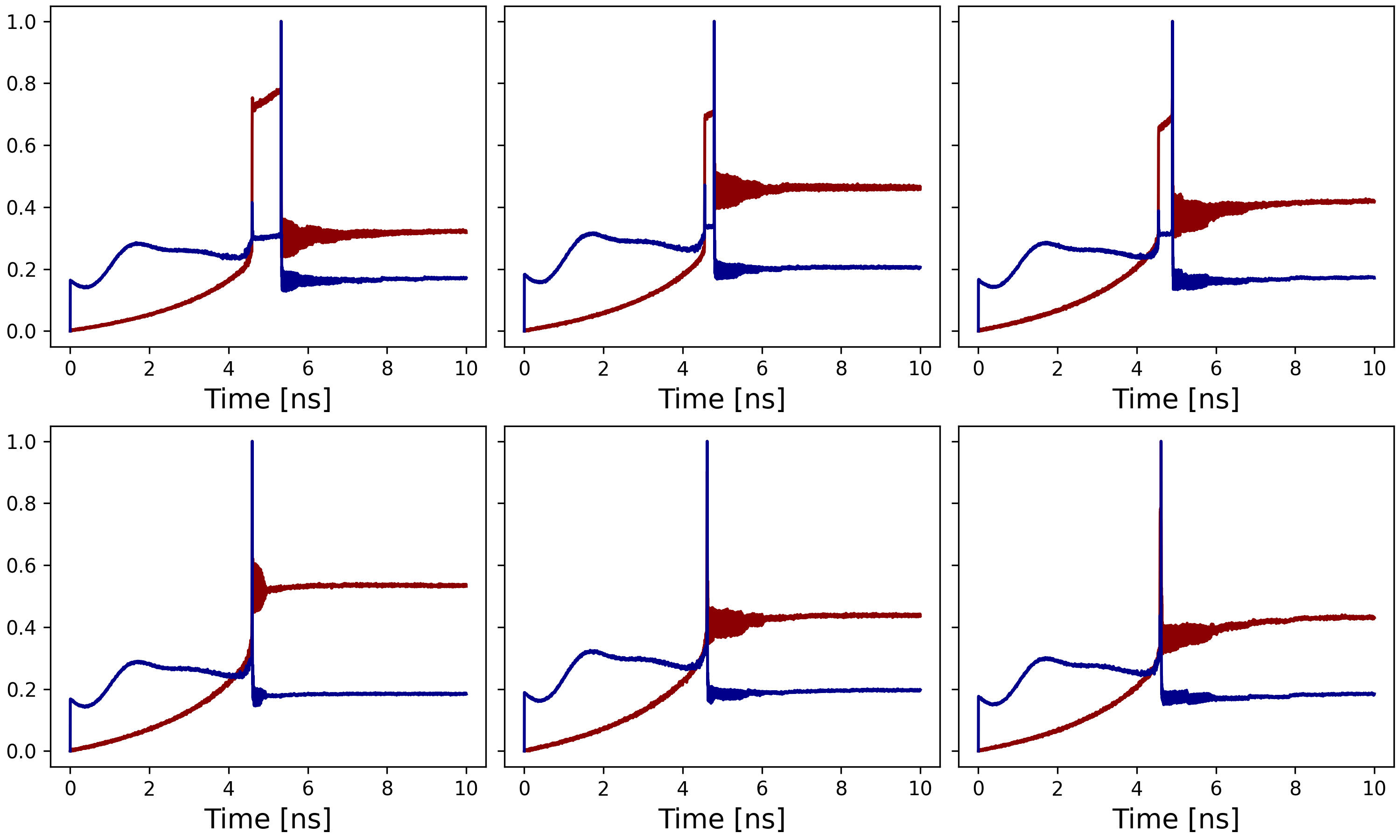}
\caption{Plots of the normalized summed LENS signal (blue lines) and summed CSP signal (red lines) as a function of time for the 6 replica simulations of the system in NPT conditions with constant applied strain rate.}
\label{fig:SUMNVT}
\end{figure}

\begin{figure}[H]
\centering\includegraphics[trim={0cm 0cm 0cm 0cm},width=\textwidth]{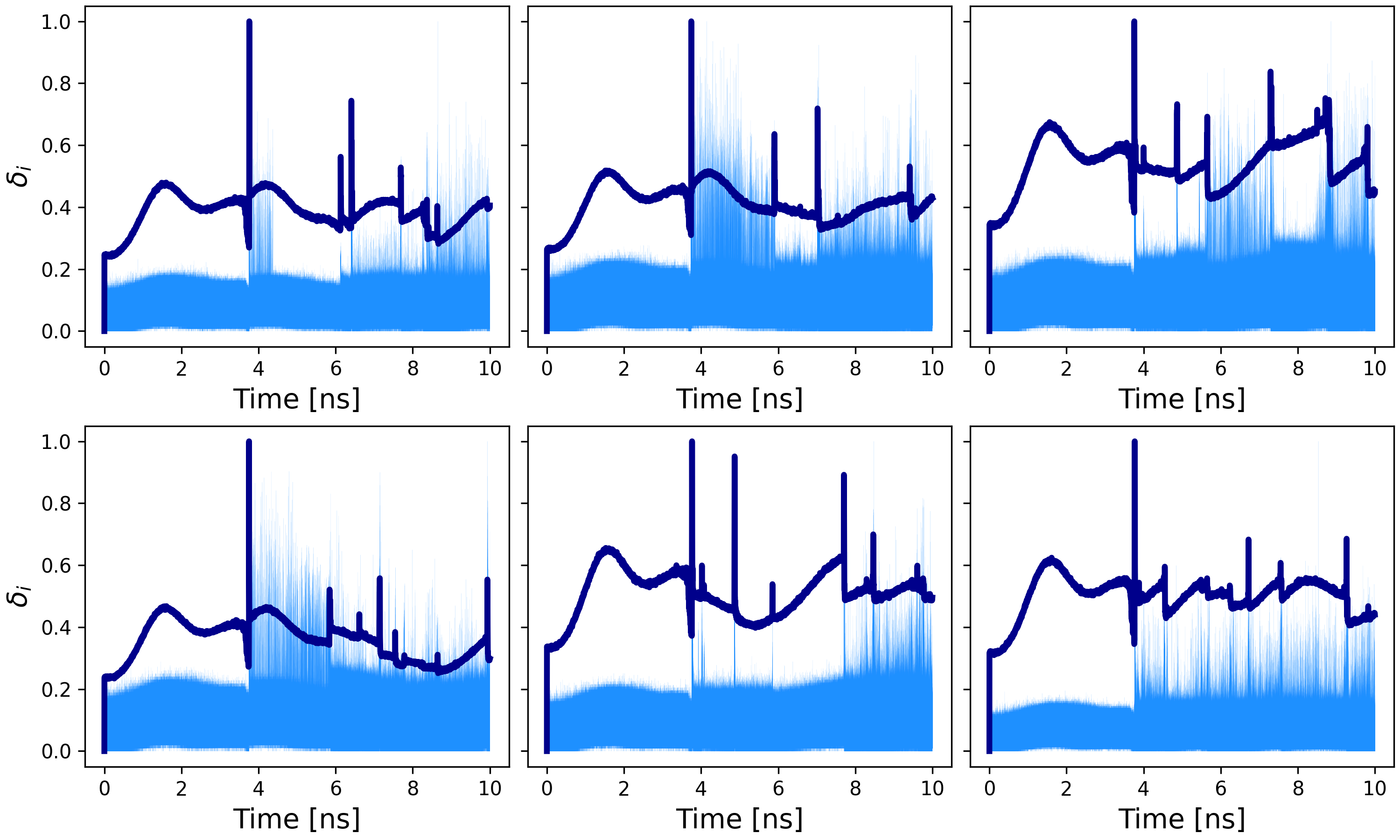}
\caption{Plots of LENS values as a function of time - light blue for the individual signals, dark blue for the summed signal - for the 6 replica simulations of the system in NPT conditions with constant applied strain rate.}
\label{fig:NPT_LENS_singolo}
\end{figure}

\begin{figure}[H]
\centering\includegraphics[trim={0cm 0cm 0cm 0cm},width=\textwidth]{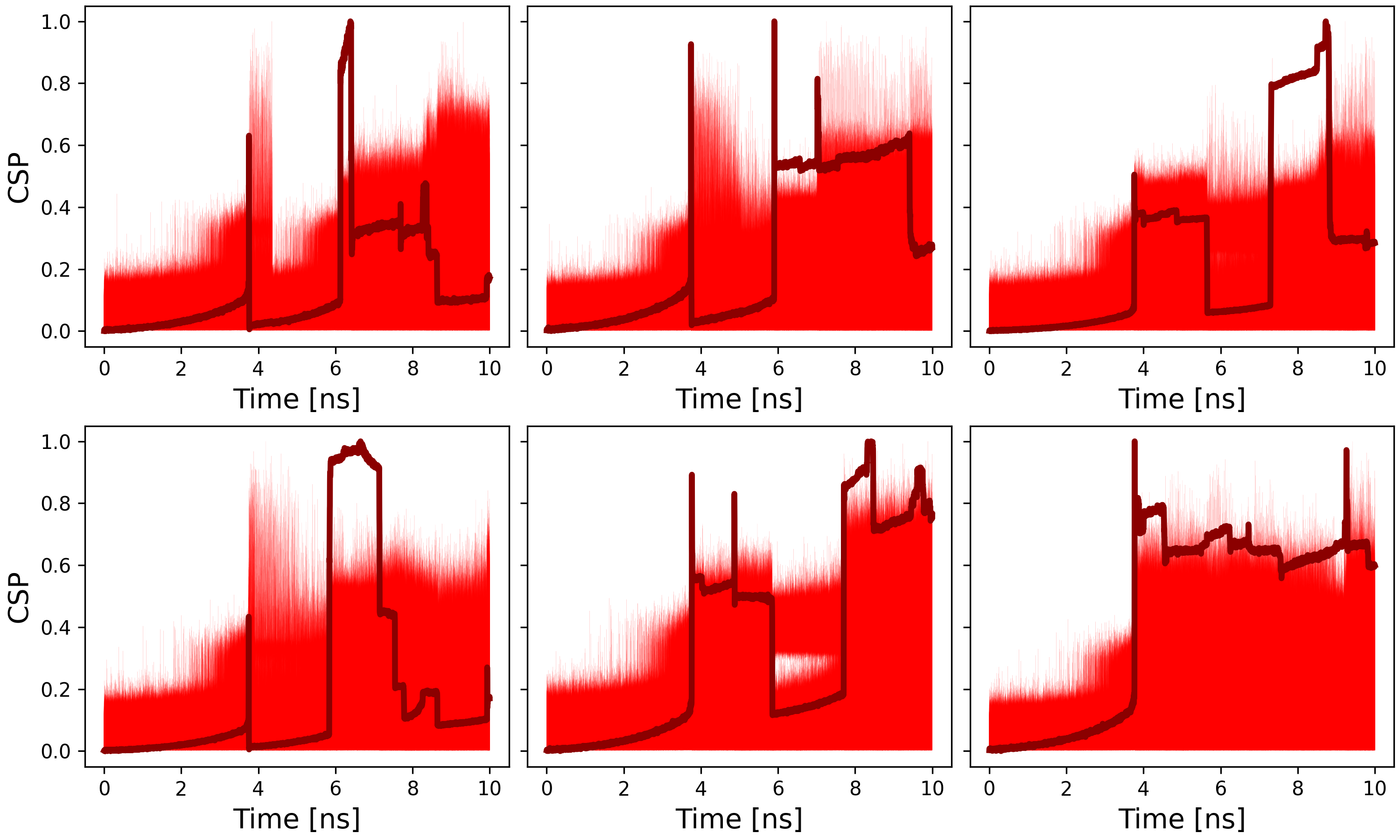}
\caption{Plots of CSP values as a function of time - red for the individual signals, dark red for the summed signal - for the 6 replica simulations of the system in NPT conditions with constant applied strain rate.}
\label{fig:NPT_CSP_singolo}
\end{figure}

\begin{figure}[H]
\centering\includegraphics[trim={0cm 0cm 0cm 0cm},width=\textwidth]{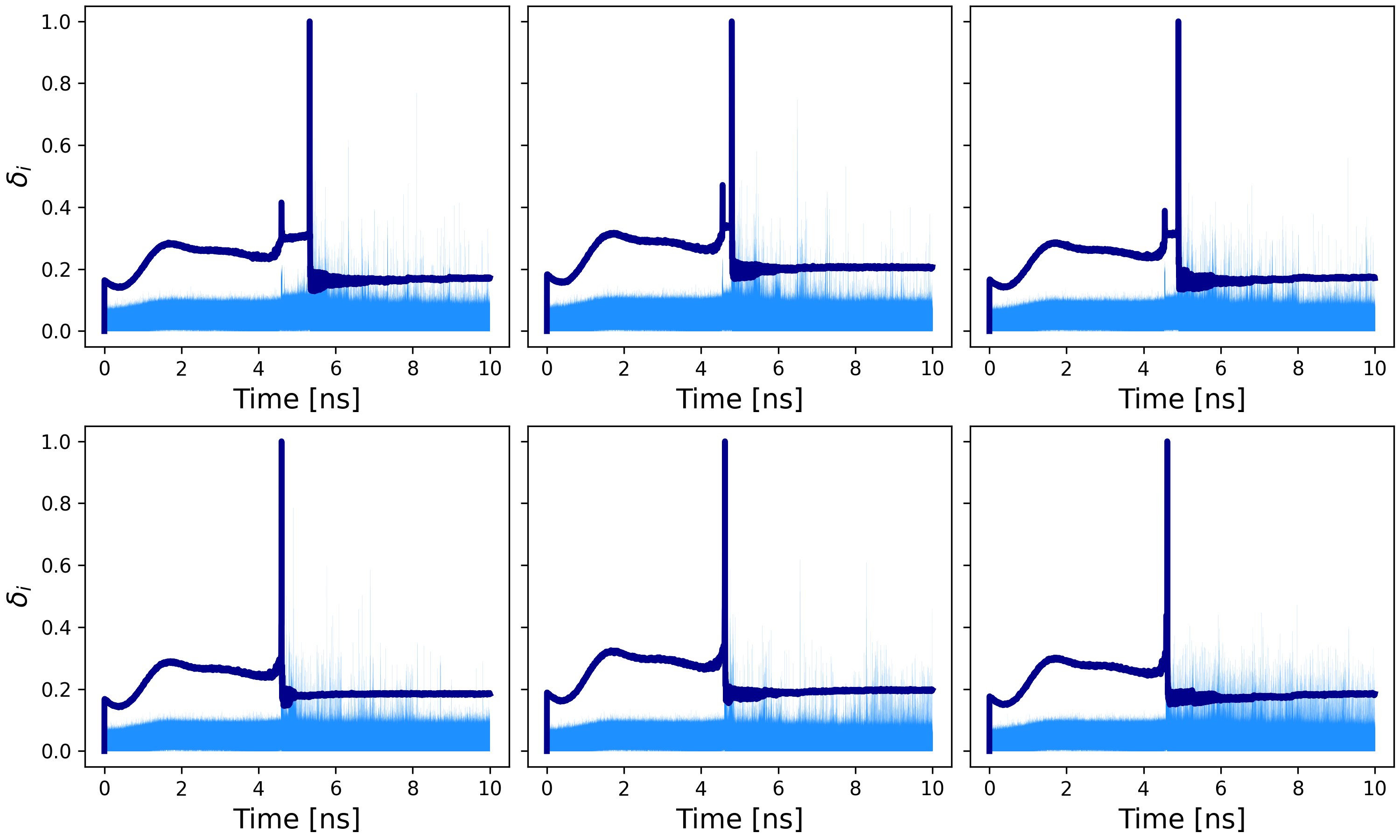}
\caption{Plots of LENS values as a function of time - light blue for the individual signals, dark blue for the summed signal - for the 6 replica simulations of the system in NVT conditions with constant applied strain rate.}
\label{fig:NVT_LENS_singolo}
\end{figure}

\begin{figure}[H]
\centering\includegraphics[trim={0cm 0cm 0cm 0cm},width=\textwidth]{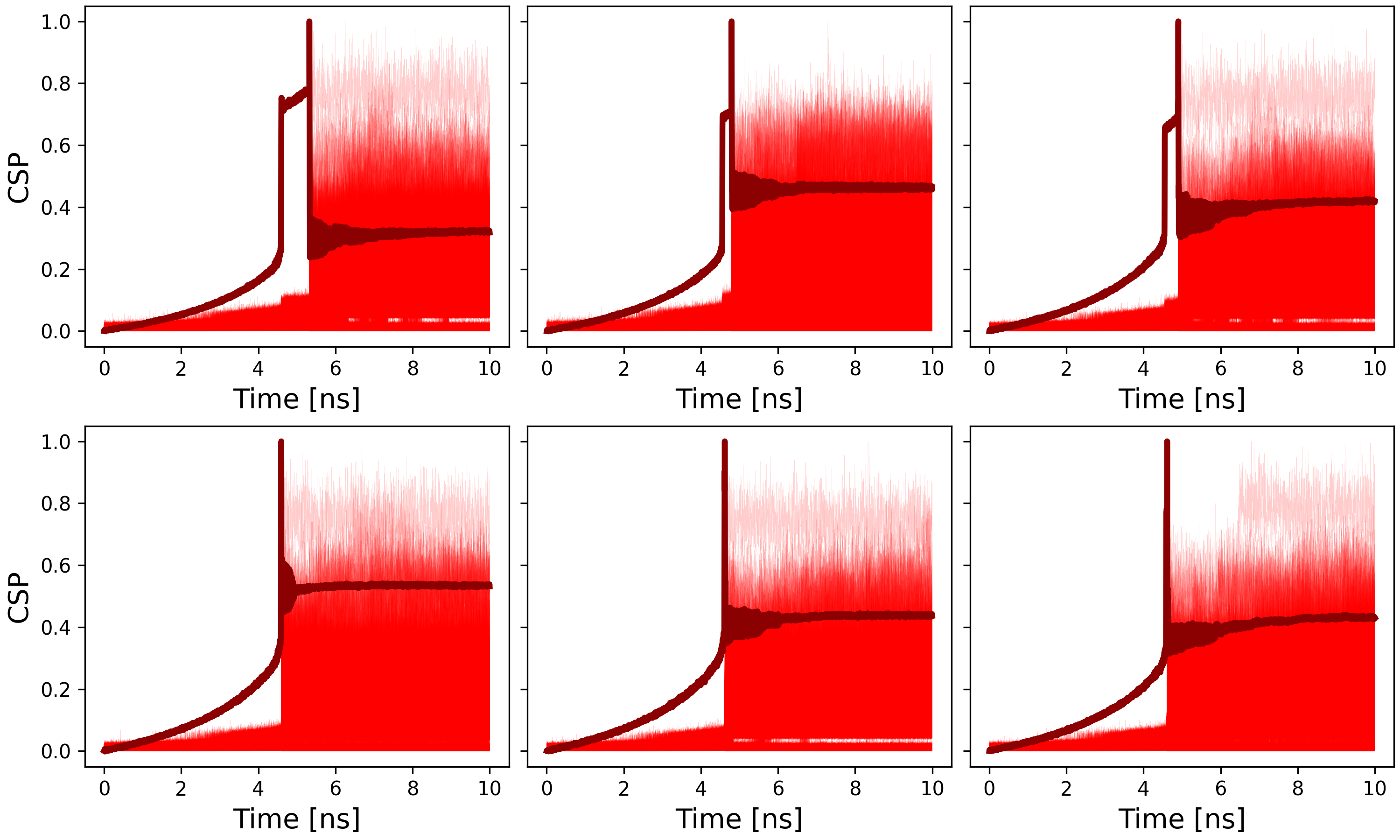}
\caption{Plots of CSP values as a function of time - red for the individual signals, dark red for the summed signal - for the 6 replica simulations of the system in NVT conditions with constant applied strain rate.}
\label{fig:NVT_CSP_singolo}
\end{figure}

\clearpage